\documentclass[graybox,secnum]{svmult}

\usepackage{amsmath,latexsym,amssymb,amsfonts,dsfont,epsfig}
\newcommand{\be}{\begin{equation}}
\newcommand{\ee}{\end{equation}}
\newcommand{\ba}{\begin{eqnarray}}
\newcommand{\ea}{\end{eqnarray}}

\usepackage{mathptmx}       % selects Times Roman as basic font
\usepackage{helvet}         % selects Helvetica as sans-serif font
\usepackage{courier}        % selects Courier as typewriter font
\usepackage{type1cm}        % activate if the above 3 fonts are
\usepackage{makeidx}         % allows index generation
\usepackage{graphicx}        % standard LaTeX graphics tool
\usepackage{multicol}        % used for the two-column index
\usepackage[bottom]{footmisc}% places footnotes at page bottom
\usepackage{hyperref}        %for hyperlinks
\usepackage{soul}            % for high-lighting of text
\hypersetup{colorlinks=true,urlcolor=blue}
\usepackage[square,numbers,sort&compress]{natbib}
\makeindex             % used for the subject index
\begin{document}
%\tableofcontents{}
\title*{Hamiltonian Theory: Dynamics}
\author{Thiemann, T. \thanks{corresponding author} and Giesel K.}
\institute{Thiemann, T. \at Inst. f. Theor. Phys. III, Dept. of Physics, 
Friedrich-Alexander Universit\"at Erlangen-N\"urnberg, Staudtstr. 7, 91058 Erlangen, Germany \email{thomas.thiemann@fau.de}
\and Giesel, K. Inst. f. Theor. Phys. III, Dept. of Physics, 
Friedrich-Alexander Universit\"at Erlangen-N\"urnberg, Staudtstr. 7, 91058 Erlangen, Germany \email{kristina.giesel@fau.de}}
\maketitle
\abstract{This chapter\footnote{This is a preprint of a chapter to appear in the 'Handbook of Quantum Gravity', edited by Cosimo Bambi, Leonardo Modesto and Ilya Shapiro, 2023, Springer, reproduced with permission of Springer.} focuses on the status of the implementation of the dynamics in the 
canonical version of Loop Quantum Gravity (LQG). Concretely this means to provide a mathematical
meaning of the quantum Einstein equations, sometimes called Wheeler-DeWitt equations,  
to give a physical interpretation and Hilbert space structure to its solutions 
and to construct a representation of the algebra of observables including a physical Hamiltonian. 
This is a structural overview intentionally skipping technical details.}

\section*{Keywords}

canonical quantum gravity, Hamiltonian constraint, Wheeler-DeWitt equations, Dirac quantisation,
reduced phase space quantisation, observables, physical Hamiltonian

\section{Introduction}
\label{s1}

The canonical or Hamiltonian approach to quantum gravity has a long tradition \cite{1a,1b,1c,1d,1e}. 
The Hamiltonian formulation of classical General Relativity (GR) is also the mathematical 
framework that underlies its initial value formulation \cite{2} in the globally hyperbolic
setting and thus is at the heart of numerical GR \cite{3} in particular in its application to 
gravitational radiation \cite{4a,4b}. A key role is played by the initial value constraints,
known as the spatial diffeomorphism $D$ and Hamiltonian constraints $C$ respectively.
These appear in the canonical formulation of any generally covariant field theory
\cite{5}. 

The physical interpretation of those constraints is as follows: In the globally hyperbolic 
setting, the spacetime manifold $M$ is foliated by a one parameter family 
of spacelike hypersurfaces $t\mapsto \Sigma_t$. The spacetime
metric $g$ can be pulled back to those, resulting in the metric $q(t)$ intrinsic to it. 
On the other hand, the extrinsic curvature $K(t)$ of those hypersurfaces provides information  
about how that intrinsic metric changes with respect to the ``time'' label $t$ of the 
foliation. Together the pair $(q(t),K(t))$ is a complete initial data set on $\Sigma_t$
coordinatising the phase space of GR. The initial value constraints are functionals 
of those and their Hamiltonian flow can be interpreted as spacetime diffeomorphisms 
tangential and transversal to the foliation. In particular, those flows together 
with the requirement that the constraints have to vanish, are in 1-1
correspondence with the Lagrangian Einstein equations. 

Although this brings canonical GR very close in appearance to the Hamiltonian 
formulation of an ordinary field theory, a conceptual difficulty is implied by the 
fact that the evolution with respect to the label $t$ is considered a gauge transformation,
namely a ``temporal diffeomorphism'' or coordinate transformation (at least 
when the equations of motion hold) which therefore is 
not observable. Moreover, GR is a field theory whose ``Hamiltonian'' is constrained
to vanish as the canonical generator of the flow is a linear combination of the 
constraints $D,C$ and observable, that is, gauge invariant quantities, are ``constants of 
motion'' with respect to this ``Hamiltonian'', i.e. they have vanishing Poisson brackets 
with all constraints when the constraints hold. This set of conceptual differences with 
respect to an ordinary field theory (say Maxwell theory on Minkowski space) is often 
called ``problem of time''. Solving these conceptual problems is therefore part of 
the answer to the question of how to address the quantum dynamics of GR.

Another peculiar aspect of those constraints is their closure, i.e. their evolution 
with respect to each other, or in other words, their stability under gauge transformations.
It turns out that the corresponding Poisson algebra of constraints has a universal 
structure \cite{5} independent of the concrete generally covariant field theory in 
question wich is called hypersurface deformation algebra $\mathfrak{h}$ and 
its exponentiation, the ``Bergmann -- Komar group'' $\mathfrak{H}$ \cite{6a,6b}.  In the strict
technical sense, $\mathfrak{h}$ is not sub Lie algebra of the Poisson algebra of 
smooth functions on the phase space (and correspondingly 
$\mathfrak{H}$ not a Lie group) but something more general that sometimes is called
an ``algebroid'' or ``open algebra''. What this means is that that the Poisson 
bracket of constraints is a linear combination of constraints, however, unlike the 
case of a true Lie algebra, the coefficients in that linear combination are non-constant
rather than constant functions on the phase space, depending on the inverse of $q$. 
In the classical theory, this has no further consequences, it just shows that 
if the initial value constraints hold on an initial hypersurface, they will hold 
on all of them, at least as long as $q$ is non-degenerate which is part of the 
definition of the globally hyperbolic setting. However, in the quantum theory one 
expects difficulties in addition to those that arise when trying to represent a classical 
Poisson Lie algebra by operators on a Hilbert space without anomalies. In particular,
as pointed out in \cite{7}, although the constraints are real valued classical
functions, they must not be represented by symmetric or even self-adjoint operators.
As the constraints encode a local gauge symmetry, anomalies are disastrous as they 
can indicate that the physical Hilbert space, which by definition is the (generalised)
kernel of the quantum constraints, encodes fewer degrees of freedom than the 
classical theory has. A proper implementation of the quantum constraints therefore 
has to address their anomaly freeness squarely. 

On the mathematical side, a major difficulty arises due to the tremendous degree of
non-linearity of GR. In an ordinary field theory the Hamiltonian is usually a 
polynomial in the canonical coordinate functions (here $(q,K)$) and the 
main problem consists in giving meaning to the part of the polynomial beyond 
quadratic order. The quadratic part typically singles out a certain Fock representation
and the higher order part, even when normal ordered, is ill-defined as products 
of operator valued distributions are involved. In fortunate cases, these ill-defined 
expressions can be tamed using perturbative renormalisation of the S matrix. 
However, in quantum gravity 
the situation is much worse because the ``Hamiltonian'' is no longer polynomial. 
Again, this cannot be avoided in a generally covariant theory because for reasons 
of gauge or general coordinate invariance only the integral of scalar densities have 
a meaning which typically involves inverse and non-integral powers of the determinant 
of the metric. This is the reason for why GR is not renormalisable \cite{8a,8b}, stated in 
Hamiltonian terms. Therefore, in quantum gravity already the proper mathematical definition 
of the objects that determine the theory presents a major difficulty which must be 
addressed before reliable physical predictions can be made. Since GR appears not to be  
perturbatively renormalisable as a standard QFT on Minkowski space in terms its perturbations 
(gravitons) of Minkowski space, it is believed that QG must be formulated non-perturbatively
and background independently.\\
\\
In this article we take a snapshot of the status of the particular incarnation of the canonical 
approach to quantum gravity coined Loop Quantum Gravity (LQG) \cite{9a,9b,9c} which is 
manifestly background independent and non-perturbative. The name 
is due to the fact that one uses a classically equivalent reformulation of GR in terms of 
connections rather than metrics and therefore one can take advantage of the 
arsenal of techniques developed for such quantum gauge theories, in particular 
gauge covariant Wilson loop variables \cite{10}. The focus will be on providing 
an overview about the different strategies that have been invented to address the 
issues mentioned above and to inform about their respective advantages, disadvantages 
and stages of development. In section \ref{s2} we begin with an overview or roadmap 
over the different developments 
of the past three decades. We try to be as non-technical as possible. There are two 
major routes, the quantum constraint approach and the reduced phase space approach which 
constrain the theory after or before quantisation respectively. These two approaches 
are typically semiclassically equivalent but may differ with respect to the quantum 
corrections and thus the corresponding phenomenology is affected by this choice. 
In section \ref{s3} we detail out the quantum constraint approach. In section \ref{s4} 
we detail out the reduced phase space approach. In section \ref{s5} we summarise and conclude.

\section{Survey of LQG dynamics}
\label{s2}

Over the past three decades a substantial amount of knowledge has been collected 
of how to address the quantum dynamics of GR in the LQG approach. Progress has been 
possible because one insisted on background independence and non-perturbative techniques.
For the first time, precise mathematical questions could be phrased and often 
answered in a concrete Hilbert space context. These developments are documented in several
100 publications with gradual improvements of each other over time resulting in different 
versions of the theory, making it difficult for the newcomer to evaluate the 
current status of the theory. The present section therefore is to serve as 
a chart.   
\begin{itemize}
\item[1.] {\it Quantisation before Constraining}
This means that one first quantises the full, unconstrained phase space and then 
imposes the constraints as operators in the corresponding representation of the 
canonical (or anti-)commutation relations (CCR or CAR) and adjointness relations (AR). In more
details this involves the following.
\begin{itemize} 
\item[1.A] ~~{\it Kinematical representations of CCR and AR}\\
In the gauge theory framework just mentioned the unconstrained phase space is 
coordinatised by a pair $(A,E)$ consisting of a real valued SU(2) connection $A$ and 
a real valued non-Abelian electric field $E$ which is the conjugate momentum of $A$.
The set of constraints $D,C$ is augmented by a Gauss constraint $G$. Then we look
for a Hilbert space representation $(\rho,\;{\cal H})$ of the CCR, schematically
$[A,A]=[E,E]=0,\; [E,A]=i\cdot 1$ and the AR, schematically $A-A^\ast=E-E^\ast=0$. To 
simplify the notation, here and in what follows
we do not distinguish between abstract algebra elements such
as $A$ and its operator representative $\rho(A)$. In quantum field theories (QFT)
the number of unitarily inequivalent representations is infinite, hence additional
physical input, usually using details of the dynamics of the QFT under consideration,
is needed to select suitable ones. In LQG one has good motivation to 
use the criterion of spatial 
diffeomorphism covariance to select an essentially unique representation 
$(\rho,{\cal H})$ \cite{11a,11b,12a,12b}
about much will be said in chapter \cite{13} of this book.
\item[1.B] ~~{\it Induced representation of the constraint algebra}\\
The challenge is then to find operator representatives of the constraints $G,D,C$ in
that selected representation. Naively, one just needs to replace in the epxression 
of the classical constraint function the variables $A,E$ by the corresponding operators.
However, this is non-trivial in several aspects. First, as 
the constraints are not linear functionals in $A,E$ an operator ordering of the 
constraints has to be chosen. Second, as $A,E$ are not operators but operator 
valued distributions on the spatial slice, we meet the problem to define the product
of such distributions common to all QFT. Third, in quantum gravity it is much worse 
than that because the constraints are not local polynomials but local 
algebraic functions of $A,E$
and thus we need to define quotients, square roots and even more singular algebraic 
functions of distributions. Fourth, all of this has to be done in such a way that the 
constraints represent $\mathfrak{h}$ without anomalies on  
a common, dense, invariant domain ${\cal D}\subset {\cal H}$ for all constraints.

A solution to the first three challenges has been given for the first time in 
\cite{13a-1a,13a-1b} both for Euclidian and Lorentzian signature, with and without 
cosmological constant, with and without standard matter. This 
was later extended to any dimension and to supergravity \cite{13b-1,13b-2,13b-3,13b-4,13b-5}. 
More precisely, one first defines regulated constraints such as $C_\epsilon$ where 
$\epsilon$ is a short distance cut-off, defines $C_\epsilon$ on the span 
of spin network functions (SNWF) $\cal D$ \cite{13c} and then takes the limit $\epsilon\to 0$
in an operator topology exploiting spatial diffeomorphism covariance and invariance thus 
arriving at continuum operators $C$ densely defined on $\cal D$. The algebra of these 
quantum constraints is non Abelian and in fact closes, but it closes with the wrong 
operator equivalents of the structure functions, although the commutator still annihilates 
spatially diffeomorphism invariant states. 

To also accomplish the fourth challenge 
in \cite{13d-1,13d-2,13d-3} it is proposed to represent $\mathfrak{h}$ not on $\cal D$ but instead 
on an invariant subspace (so called ``habitat'' \cite{13e}) 
of ${\cal D}^\ast$, the space of algebraic (i.e. discontinuous) distributions on $\cal D$, 
at the price of changing the density weight of $C$. At the moment, this approach 
is geared to the Euclidian signature vacuum theory because one exploits the fact 
that in this case $C$ very much looks like $D$ but with an electric field dependent 
vector field as generator of one parameter families of diffeomorphisms (so called 
``electric shift approach''). Another approach to the fourth challenge is (Hamiltonian)
renormalisation \cite{13f} which is complementary to 
the renormalisation \cite{14a-1,14a-2} in the covariant (spin foam) approach to LQG. 
As pointed out in \cite{13g}, the very definition of 
$\mathfrak{h}$ in the classical theory relies on the assumption that the metric $q$ 
be regular. On the other hand, the quantum metric annihilates the dense domain 
$\cal D$ almost everywhere. The difficulties to accomplish the fourth challenge 
turn out to be tightly connected with the issue of quantum non-degeneracy and it 
is therefore suggested to make quantum non degeneracy an integral part of notion 
of anomaly freeness. By construction, the renomalisation programme systematically builds 
such non degenerate representations of the CCR and AR. Yet another approach to the 
fourth challenge is the master constraint proposal, see immediately below.
\item[1.C] ~~{\it Kernel of the quantum constraints}\\
Assuming that all four challenges of the previous task have been met, the next step 
is to solve the constraints and to pass from 
the unphysical or kinematical Hilbert space $\cal H$ to the physical Hilbert space
${\cal H}_{{\sf phys}}$. Unless zero is only in the pure point spectrum of all constraints,
the physical Hilbert space is not a subspace of the kinematical one, thus the formal
solutions $\psi$ to the constraint equations $G\psi=D\psi=C\psi=0$ must be equipped 
with a new scalar product. If the constraints would form a Lie algebra of self-adjoint 
operators
then one could use rigging or group averaging techniques \cite{14} which essentially 
is the map $\eta \psi=\delta(G,D,C)\psi$ where the $\delta$ 
distribution is obtained by passing from $\mathfrak{h}$ to the Lie group 
$\mathfrak{H}$ and integrating the unitary operators corresponding to the constraints
over the Lie group. The new inner product is then essentially 
$<\eta \psi, \eta \psi'>_{{\sf phys}}=<\psi,\eta \psi'>$.

As this is not the case, one has two options. The first is the master constraint approach  
\cite{15}, the second is to pass to classically to equivalent constraints (typically 
Abelian) that do form a Lie algebra \cite{16}. 

As far as the master constraint is 
concerned, the observation is that instead of considering the individual constraints
one can equivalently construct their weighted sum (or rather integral) $M$ of squares
called master constraint. The classical constraint surface of the phase space 
is then encoded by the single equation ${\bf M}=0$ and a classical observable $O$ is specified 
the single equation $\{\{{\bf M},O\},O\}_{{\bf M}=0}=0$. In the quantum theory one then correspondingly 
also has to impose only a single equation $M\psi=0$ which can be solved using 
rigging or direct integral decomposition (DID) \cite{17} techniques. An attractive 
feature of this approach is that, since there is only one constraint left, one no
longer has to worry about the representation of 
$\mathfrak{h}$ and its possible anomalies. However, while the spectrum of $\bf M$ is 
granted to be a subset of the non-negative reals, zero may not be included and 
therefore one may need to subtract from $M$ a constant to allow for sufficiently 
many solutions. That constant is therefore a manifestation of the anomaly of $\mathfrak{h}$
in the master constraint approach. 

As far as constraint
Abelianisation is concerned, the idea is to solve the constraints, collectively called 
$Z$, for certain momenta $p$ i.e. to write them in the equivalent form 
$\tilde{Z}=p+h$ where $h$ does not depend on $p$. Then one can show that the $\tilde{Z}$
have vanishing Poisson brackets among themselves \cite{18} and therefore rigging techniques
can be used in principle if one is able to find a representation of that Abelian algebra.     
\item[1.D] ~~{\it Induced representation of the quantum observables}\\ 
Finally, supposing that the physical Hilbert space has been constructed we still 
have not done any physics with it. Abstract quantum observables are the 
self-adjoint operators $O$ defined densely on ${\cal H}_{{\sf phys}}$. Thus they 
preserve the joint kernel of the constraints and thus
have vanishing commutators with all constraints on that kernel. However, a priori they lack 
any physical interpretation and thus useful observables will be representations 
of classical observables (vanishing Poisson brackets with all constraints when the 
constraints hold). This will again
meet the same mathematical challenges as the constraints themselves because the 
observables of GR are expected to be non-polynomial (and in case of vacuum GR even 
non-local \cite{19}). Also, one has to extract a non-trivial physical time evolution (or 
physical Hamiltonian) for those observables which reduces to the time evolution of 
(standard) matter observables when fluctuations of GR around exact vacuum solutions 
(e.g. Minkowski) are negligible.  
\end{itemize}
\item[2.] {\it Quantisation after Constraining}\\
In the reduced phase space approach one aims at quantising the physical degrees of freedom only. For this purpose one solves the constraints already at the classical level by means of constructing classical observables and takes their corresponding algebra as the starting point for the quantisation of the CCR and AR. The dynamics of these observables is generated by a so called physical Hamiltonian that will be implemented as a corresponding operator directly on the physical Hilbert space. Therefore, non-observable quantities are never subject to quantisation, anomalies cannot arise. Given the severe difficulties of the quantisation before constraining approach
sketched above, the reduced phase space approach appears to be especially attractive
and much more economic. However, whether the quantisation programme can be completed along these lines crucially depends on finding a representation of the CCR and AR of these
observables because in general these algebras have a more complicated structure than their kinematical counter parts of the non observables $A,E$. Those observables can be explicitly constructed in the framework of the relational formalism \cite{22a,22b,22c,22d,22e}. A key ingredient is to choose some reference fields as "clocks" and "rods" as a physical dynamical reference system with respect to which the observables and dynamics of the remaining degrees of freedom are formulated. What kind of reference fields are chosen here is free at the first place and determines the reduced model as the reduced phase space as well as the physical Hamiltonian depend on these choices. This is of course expected, even on Minkowski space the Hamiltonians of a free Klein-Gordon field depend on the
inertial frame that one uses, differing by a corresponding boost generator.   Often one distinguishes between so called geometrical and matter reference fields that are chosen from the corresponding sectors of the theory but of course a mixture will also be possible. In full vacuum GR such a manageable $^\ast-$algebra has not been found to date. So far those geometrical clocks have only been considered on a perturbative level \cite{GeomClocks-1,GeomClocks-2,GeomClocks-3,GeomClocks-4} which is why most efforts in that direction use matter in a crucial way.

\begin{itemize}
\item[2.A.] ~~{\it Choice and phenomenology of material reference systems}\\
Working with physical dynamical reference systems takes into account that in realistic models something like an idealised test observer that does not back react on the system does not exist and can at most only be assumed in some limit of the models. In order to serve as a system of ``rods and clocks'' the matter must be present ``everywhere and every time''. In other words, it must have non-vanishing energy momentum density throughout spacetime and considered as a map from $M$ to itself it should be a diffeomorphism. Given a system to start that couples matter dynamically to gravity one chooses in general a suitable subset of the matter degrees as reference fields and then constructs the reduced theory of the remaining physical observables. Choosing a different subset as matter reference fields yields to another reduced theory with the same amount of physical degrees of freedom and at the classical level the relation among these two reduced models is well established. Not any chosen matter type will necessarily yield to a manageable observable algebra for which the quantisation programme can be completed and therefore it is often convenient to choose a reference matter model for which we can mimic in some limit the idealised test observer because it influences the "observed matter and geometry " as little as possible.  A matter species that appears to come close to have those properties
is dark matter DM \cite{20} for it interacts only gravitationally and seems to be 
omnipresent throughout the universe, although presently it is unclear what DM actually is.  
In the LQG literature several matter types were employed, mostly driven by mathematical 
convenience and these matter types were coupled in addition to standard model matter. Typically these are scalar fields with an energy momentum tensor of perfect fluid form,
with dust matter being not only particularly convenient but also because it comes closest
to the notion of a test observer that only interacts gravitationally \cite{21-1,21-2}. For instance in the case of cosmology this means one works with 2-fluid models instead of 1-fluid models \cite{LQCclocks}. Since by construction such models carry more physical degrees of freedom than the corresponding models without the reference matter, one needs to carefully analyse the phenomenological implication of these reduced models, thereby possibly constraining the (dark) matter model employed. This has been done e.g. in 
\cite{21a-1,21a-2} where it is shown that for the Einstein-inflaton system with four additional dust fields the additional polarisations here assigned to the gravitational sector rapidly decay in the late universe. Furthermore, in \cite{LQCclocks} the effect of different choices of clocks  on inflation was investigated showing that there exist initial conditions for which dust as well as scalar field clocks behave as "good" clocks that influence the observed system only in a tiny way.

\item[2.B] ~~{\it Deriving the reduced phase space}\\
Once suitable reference matter the reduced phase space is obtained by rewriting the set of constraints in an equivalent form that is convenient for the construction of the observables.
\\
{\it Abelianisation of the constraints:}
\\
For each constraint one reference matter field is chosen and this field needs to be, at least weakly canonically conjugate to the corresponding constraint. In the matter models considered in \cite{21-1,21-2,23a,23b,23c,23d,23e} this can be obtained by solving the constraints, denoted by $Z_I$, for the reference matter momenta $\pi_I$ conjugate to the material reference matter $\phi^I$ where $I$ labels the set of constraints. A physically equivalent form of the constraints is then given by $\tilde{Z_I}=\pi_I+h_I$ where $h_I$ depends on all variables of the unconstrained phase space except the $\pi_I$. This set of constraints is then by construction Abelian using that the original constraints are first class. While in principle such a (weak) Abelianisation is always possible, at least locally, it usually requires to solve PDE's to obtain the $\tilde{Z}_I$ if the momenta $\pi_I$ are not involved only algebraically in the $Z_I$. The latter is the case for the gravitational degrees of freedom and explains why using matter is practically important. As the standard dependence in the Hamiltonian constraint on $\pi_I$ is typically quadratically, a drawback is that the phase space decomposes into branches related to the sign ambiguties of $h$. As a consequence, $Z_I,\;\tilde{Z}_I$ are strictly equivalent only on one of those branches. This may be
considered a mild price to pay in view of the huge amount of advantages that one 
otherwise gains over the quantisation before constraining method.
~\\
{\it Construction of observables and physical Hamiltonians:}
~\\
To construct observables one introduces  as set of gauge fixing conditions,  denoted by $F_I=
\phi^I-\tau^I$ where $\tau^I$ are fixed functions 
on the spacetime manifold $M$ sometimes called 'multi-fingered' times. The constraints written in the form $\tilde{Z}_I$ have the property that their corresponding 'Dirac matrix' $M^I_{\,\,}:=\{F^J,\tilde{Z}_J\}$ has a non-vanishing determinant being the requirement admissable gauge fixing conditions need to satisfy. Denoting the remaining conjugate pair, sometimes called 'true degrees of freedom', collectively by $(Q,P)$, suppressing any index structure here, one can construct 
 a "relational" observable \cite{22a,22b,22c,22d,22e} $O_K$ associated to a generic phase space functional $K$ of $(Q,P)$ as 
$
O_K(\tau):=\big[\exp(u X_{\tilde{Z}})\; \cdot\; K\big]\big|_{u=F}
$
with $X_{\tilde{Z}}$ being the Hamiltonian vector field of the constraint $\tilde{Z}$. Its interpretation is that it returns the value of $K$ when $\phi$ takes the value $\tau$ underlying their relational nature. In particular $O_K$ agrees with $K$ in the gauge $F=0$ and thus can be understood as the corresponding gauge invariant extension of $K$.  The so obtained $O_K$'s have vanishing Poisson brackets with all constraints $\tilde{Z}$ and are thus gauge invariant.  Since $K,K'$
depend only on $Q,P$ we have $\{O_K,\;O_{K'}\}=O_{\{K,K'\}}$ for the equal 
$\tau$ brackets, i.e. the CCR and AR of 
$K,\;K'$ and $O_K,\; O_{K'}$ respectively are isomorphic. Consequently, if one considers $Q,P$ one can conclude that $O_Q,O_P$ remain canonically conjugate and thus retain simple CCR and AR meaning that quantising the observable algebra is not more difficult than the corresponding kinematical algebra. The dynamics of the $O_K$ can be understood as a physical evolution with respect to the parameter $\tau_0$ involved in gauge fixing condition $F_0=\phi_0-\tau_0$ associated the Hamiltonian constraint $C=\pi_0+h_0$, where the label zero was introduced to label the corresponding quantities for the Hamiltonian constraint. The generator of the dynamics, the physical Hamiltonian, is given by the observable $H:=O_{h_0}$ that is a functional of the observables $O_Q,O_P$ only and can in principle depend on $\tau_0$ and thus be a time-dependent Hamiltonian. Its explicit form depends on the chosen reduced model and specific examples will be discussed in section \ref{s4}.

\item[2.C] ~~{\it Quantisation of the reduced phase}\\
The final step consists in picking a representation 
$(\rho_{{\sf phys}},\; {\cal H}_{{\sf phys}})$ of the CCR and AR of $(Q,P)$. Again 
this needs physical input. As we are now in the situation of an ordinary 
Hamiltonian system it is natural to impose the condition that $H$ when expressed 
in terms of the operators corresponding to $Q,P$ is promoted to a densely defined 
self-adjoint operator on ${\cal H}_{{\sf phys}}$ after suitable regularisation and 
renormalisation steps. Since one can choose $(Q,P)$ to be $(A,E)$ one can pick 
the same representation as in the quantisation before constraining route just that 
now the Hilbert space is the physical Hilbert space and not the kinematical one.

Defining the Hamiltonian in that representation meets the same mathematical challenges as 
for defining the constraints themselves, depending on the matter employed, 
these are even worse because of the 
additional square roots involved in the expression for $h$ which is an algebraic function 
of $q, G, D, C$. Nevertheless, in \cite{23a,23b,23c,23d,23e} it was shown that 
using the same methods as in \cite{13a-1a,13a-1b} one can arrive at a positive operator $H$ on 
${\cal H}_{{\sf phys}}$ which therefore has at least one self-adjoint (Friedrichs) extension. 
Moreover, the operator $H$, just as its classical analog, is Gauss and spatially 
diffeomorphism invariant (with respect to active rather than passive diffeomorphisms)
and therefore free of part of the possible anomalies. 

A crucial difference between the action of $H$ and the constraints is the following:
The Hilbert space is a direct sum of mutually orthogonal subspaces which are labelled
by piecewise analytic (more precisely semi-analytic, see \cite{13}) graphs. Thus the 
Hilbert space is not separable. Due to this, the fact that the (active) spatial diffeomorphisms 
mix those sectors and since $H$ is spatially diffeomorphism invariant, the operator 
$H$ must not mix those sectors, otherwise it could not be densely defined. This feature 
has been coined ``non-graph changing'' and applies to any spatially diffeomorphism 
invariant operator. By contrast, the constraint operators are not spatially diffeomorphism 
invariant but just co-variant, thus their action does mix those sectors and are thus 
called ``graph changing''. This difference has important consequences for the semiclassical
limit with respect to coherent states \cite{24a,24b,24c,24d} on fixed graphs: While the semiclassical 
limit of $H$ coincides with the classical expression plus quantum corrections for sufficiently
fine and large graphs, the semiclassical limit of the constraints with respect to the 
same states does not yield the expected result. Presently no semiclassical states are 
known with respect to which the constraints have an appropriate limit but superpositions 
of the coherent states over those graphs generated by the constraints appear to be 
promising candidates.

~\\
~\\
The fact that $H$ must be non-graph changing and appears to have good semiclassical behaviour 
on sufficiently large and fine graphs has motivated the Algebraic Quantum Gravity (AQG)
version of LQG \cite{25a,25b,25c,25d}. Here one considers a single, infinite abstract (algebraic) graph 
once and for all. The term ``algebraic'' means that the graph is completely specified 
by its adjacency matrix encoding just the information which vertices are connected by which 
edges. All quantum degrees of freedom and the Hilbert space 
refer to those abstract edges and vertices only, 
thus defining an abstract $^\ast-$algebra of operators. Also the Hamiltonian $H$ is 
expressed in terms of these algebra elements and it now acts on the chosen abstract graph
which it preserves by construction and thus is non-graph changing. However, it is not 
required to be sub-graph preserving and in fact does not. 
The information of how the abstract graph is embedded into the spatial hypersurface 
(knotting and braiding of edges) is supplied by the semiclassical state with 
respect to which the semiclassical limit is under good control. If the abstract 
graph is infinite one can choose that embedding arbitrarily densely so that in this
sense a continuum limit is partially obtained. 

One additional advantage 
of AQG over LQG is that it reduces the degree of non-separability of the physical Hilbert 
space: In the case of compact spatial slices it becomes separable, in the non-compact
case it is still non-separable but in a more controlled fashion involving the so-called
infinite tensor product (ITP) \cite{26}. 
The non-separability of the (physical) Hilbert space is anyway a source of several 
complications such as the discontinuous action of spatial diffeomorphisms and 
holonomy operators, the fact that no SNWF function exist which are excited everywhere
on the spatial slice and which therefore describe a degenerate quantum geometry. 
It is also a vast over-coordinatisation of the degrees of freedom as e.g. much fewer 
graphs would suffice to separate the points in the space of classical connections.         
AQG therefore can be considered as the starting point for the Hamiltonian renormalisation
approach to LQG \cite{13f} where in addition one tries to remove the dependence on 
the abstract graph of AQG thereby fully reaching the continuum limit.   

~\\
Another advantage of AQG over LQG is the issue of propagation \cite{27}: In 
order that the commutator of Hamiltonian constraints annihilates spatially diffeomorphism
invariant distributions, the constraints defined in \cite{13a-1a,13a-1b,13a-2} 
are such that a second 
action does not react to the changes made to the graph underlying a SNWF that were 
done by the first action. This appears to induce an unphysical ultra-locality 
and thus absence of propagation in the sense that the Hamiltonian constraint action 
can be described in terms of its action at individual vertices that do not influence 
each other. That this is not the case beyond reasonable doubt has been demonstrated 
in \cite{28} where 
it is pointed out that communication between vertices does happen at the level 
of the solutions to all constraints. The basic mechanism is that a solution is also 
spatially diffeomorphism invariant and therefore the constraint information coming from 
its action at adjacent vertices cannot be uniquely assigned to one or the other of those
vertices any more, i.e. the information spreads out. In AQG this propagation issue 
does not arise from the outset because the Hamiltonian is not sub-graph preserving. 
\end{itemize}
\item[3.] {\it Connection with covariant quantisation -- path integrals}\\
The precise correspondence between canonical and covariant quantisation of constrained 
systems, no matter how complicated the constraints and their algebra may be, 
is by now well understood and subject of many excellent textbooks such as
\cite{18}. Basically it starts from the reduced phase space description above which 
is an ordinary Hamiltonian system. Its phase space path integral description of 
inner products between physical states $\Psi,\;\Psi'$ can therefore be 
obtained by standard methods of (Euclidian) QFT. Then one unfolds the reduced phase space path integral into a 
path integral over the full 
kinematical phase space using appropriate $\delta$ distributions and measure factors 
(essentially the square root of the modulus of the determinant of the Poisson bracket 
matrix between 
second class constraints and the 
modulus of the determinant of the Poisson bracket matrix between
the first class constraints and the gauge 
fixing conditions). The application of this general framework to GR in the gauge theory 
description has been described in \cite{16}. One can view the resulting formula 
as an explicit implementation of the rigging map $\Psi=\eta \psi,\; \Psi'=\eta \psi'$ 
for purely first class constrained 
systems mentioned above with $\eta=\delta[\tilde{Z}]$. 

It is important to stress that 
here we must use the Abelianised version $\tilde{Z}$ of the constraints $Z=(G,D,C)$ 
as otherwise group averaging cannot be performed due to the fact that the algebra of 
the $Z$ is not a Lie algebra. Note that this can be done for arbitrary (standard) matter 
coupling. The final path integral is over the unconstrained phase space. Only in simple 
cases can one carry out the momentum integrals and arrive at a path integral just 
over the kinematical configuration space.   

In the spin foam approach to LQG described in chapters \cite{29a,29b} of this book
one chooses a different starting point. Namely one postulates the rigging inner 
product as the 
path integral over $A, B, C$ of the exponential of i times 
the Plebanski action and the boundary states
$\psi, \psi'$ which depend only on the Lorentz connection $A$ (for the chosen 
signature). The Plebanski action contains
a $B\wedge F$ term where $F$ is the curvature of $A$ and $B$ is a Lie algebra valued two 
form and a term of the form $C\cdot B\wedge B$ where $C$ is a Lagrange multiplier imposing 
certain simplicity constraints on $B$ ensuring that $B$ is derived from a tetrad (so 
that the BF terms becomes the Palatini action). To date, establishing
the precise relation between 
the Plebanski version and the reduced phase derivation of the covariant formulation
sketched above has not been achieved.  
\end{itemize}

\section{Constraining after quantisation}
\label{s3}

This section goes into more details with respect to the Dirac quantisation approach 
of LQG. 
%We focus on concepts and skip over rather technical details which would distract
%attention from the more important main ideas. Readers interested in those details are 
5encouraged to study the original literature that we refer to along the presentation.
We will consider only 4 spacetime dimensions and standard matter, see 
\cite{13b-1,13b-2,13b-3,13b-4,13b-5} for how to pass beyond those limitations.

\subsection{Notation}
\label{s3.1}

~\\
We begin with the notation:\\
\\
$a,b,c,.. =1,..,3$: spatial tensor indices\\
$A,B,C,.. =1,2$: chiral Fermion (spinorial) index\\
$\epsilon_{AB},\;\epsilon^{AB}$: completely skew spinor metric\\
$j,k,l,.. =1,..,3$: su(2) Lie algebra index\\
$J,K,L,.. =1,..,d$: Lie algebra index of compact gauge group $G$ of dimension d\\
$\iota,\kappa,\lambda = 1,..,N$: defining represention index of compact gauge gropup G\\
$\alpha,\beta,\gamma = 0,..,n-1$: dark matter or dust species label\\
$\delta_{jk},\;\delta^{jk}$: Kronecker symbol, 
similar $\delta_{JK},\;\delta_{\iota\kappa},\delta_{\alpha\beta},\delta^{\alpha\beta}$\\
$A_a^j$: gravitational connection\\
$E^a_j$: gravitational electric field; momentum conjugate to $A$\\
$\Gamma_a^j$: spin connection determined by $E$\\
$\Gamma^a_{bc}$: Levi-Civita connection determined by $E$\\
$\underline{A}_a^J$: Yang-Mills connection\\
$\underline{E}^a_J$: Yang-Mills electric field; momentum conjugate to $\underline{A}$\\
$\eta^A_\iota$: chiral Fermion field with YM charge\\
$\nu^A$: chiral fermion field without YM charge (neutrino singlett)\\    
$\phi_\iota$: Higgs field for $G$ (generically complex valued)\\
$\pi^\iota$: conjugate momentum of Higgs field\\
$X^\alpha$: dark matter or dust scalars \\
$Y_\alpha$: conjugate momenta of dark matter or dust fields\\
$\Lambda$: cosmological constant\\
$T_j,\underline{T}_J$: respective Lie algebra basis.\\
$\tau_j,\underline{\tau}_J$: respective Lie algebra basis matrix in defining representation.\\
$\nabla$: gauge co-variant differential annihilating $E$\\
${\cal D}$: gauge co-variant differential replacing $\Gamma_a^j$ by $A_a^j$\\
\\
We have chosen Lie algebra basis elements such that 
$Tr(\tau_j\tau_k)=\delta_{jk},\;{\rm Tr}(\underline{\tau}_J\underline{\tau}_K)=\delta_{JK}$
and thus do not need to pay attention to the position of these indices.
If $G$ is zero- or one-dimensional we simply drop the index $\iota$ allowing to treat hypothetical 
scalar fields such as the inflaton. The differentials $\nabla, {\cal D}$ 
act on tensorial, spinorial ,
representation and 
Lie algebra indices, i.e. for a hypothetical field $X^{aA}\;_{j\iota J}$ of density 
weight zero
\be \label{3.1}
\nabla_b X^{aA}_{j\iota J}=\partial_b X^{aA}_{j\iota J}+\Gamma^a_{bc}\; X^{cA}_{j\iota J}
+\Gamma_b^k\; [T_k]_j\;^l\; X^{aA}_{l\iota J}
+\Gamma_b^k\; [\tau_k]^A\;_B\; X^{aB}_{j\iota J} +
\underline{A}_b^K\; [\underline{\tau}_K]_\iota\;^\lambda\; X^{aA}_{j\lambda J}+
\underline{A}_b^K\; [\underline{T}_K]_J\;^L\; X^{aA}_{j\iota L}
\ee
We are considering only one matter species of each kind, the case of several species possibly
for 
different gauge groups can be obtained by simply adding those terms. In particular
fermions of opposite chirality can be treated by interchanging $\eta^A_\iota$ with its adjoint
$(\eta^A_\iota)^\ast$ which up to a factor of $i$ plays the role its conjugate momentum.
Majorana fermions also can be treated like this, instead of having two independent 
chiralities $\eta, (\eta')^\ast$ we have $\eta'=\eta$.\\
\\
From the quantum theory of detailed in chapter \cite{13} we only need to know that there 
is a Hilbert space $\cal H$ with dense subset $\cal D$ consisting of the span of the 
orthonormal spin network function (SNWF) basis. A SNWF $T_{\gamma,j,\iota}$ is labelled by 
a triple $\gamma,j,\iota$ where $\gamma$ is a piecewise (more precisely semi-) analytic, oriented 
graph with edge set $E(\gamma)$ and vertex set $V(\gamma)$, $j$ is a set of irreducible 
representations $j_e$ (spins and YM irreducibles for SU(2) x G)
colouring the edges $e\in E(\gamma)$ and $\iota$ is a set of combinations 
of irreducible representations and gauge invariant intertwiners 
$\iota_v$ colouring the vertices $v\in V(\gamma)$. The fields $A,\underline{A}$ are excited 
along the edges, the fields $\eta,\nu,\phi,X$ on the vertices. 

\subsection{Lagrangian, Legendre transform, constraints}
\label{s3.2}

The Lagrangian for GR coupled to matter is a spacetime scalar density of weight one in order 
that the corresponding action is spacetime diffeomorphism invariant. Therefore the Legendre 
transform with respect to a foliation $t\mapsto \Sigma_t$ of $M$ yields constraints 
which are tensor densities of weight one. Thus {\it density weight one is the natural
density weight which is dictated by general covariance}. The leaves $\Sigma_t$ 
of the foliation are mutually spatially diffeomorphic by global hyperbolicity and thus 
diffeomorphic to some given 3-manifold $\sigma$. The fields listed in the previous 
subsection can be considered as fields on the manifold $\mathbb{R}\times \sigma$ and 
after performing the Legendre transform the initial value constraints take the following 
form:
\ba \label{3.3}
G_j &=& \partial_a E^a_j+\epsilon_{jkl} A_a^k E^a_l+J_j
\nonumber\\
\underline{G}_J &=& \partial_a \underline{E}^a_J+f_{JKL} \underline{A}_a^K 
\underline{E}^a_L+\underline{J}_J + \pi^T\underline{\tau}_J \phi
\nonumber\\
D_a &=& E^b_j 
\partial_a A_b^j-\partial_b(E^b_j A_a^j)
+\underline{E}^b_J\partial_a \underline{A}_b^J-\partial_b(\underline{E}^b_J \underline{A}_a^J)
\nonumber\\
&& +i\delta_{AB}[(\eta^A_\iota)^\ast\;\delta^{\iota\kappa}\;\partial_a\eta^B_\kappa
+(\nu^A)^\ast\;\partial_a \nu^B-c.c.]
+\pi^\iota\;\partial_a \phi_\iota + Y_\alpha\;\partial_a X^\alpha
\nonumber\\
C &=& C^G_E+C^G_L+C^C+C^{YM}+C^F+C^S+C^Y+C^D
\nonumber\\
C^G_E &=& \frac{F_{ab}^j\epsilon_{jkl} E^a_k E^b_l}{|\det(E)|^{1/2}},\;\;\;
C^G_L = 2\beta \;\frac{K_a^j K_b^l E^a_{[j} E^b_{l]}}{|\det(E)|^{1/2}},\;\;\;
C^C  =  \Lambda \;|\det(E)|^{1/2}
\nonumber\\
C^{YM} &=& \frac{1}{2}\; E_a^j E_b^k\;\delta_{jk}\; |\det(E)|^{1/2}\;[\underline{E}^a_J 
\underline{E}^b_K+\frac{1}{2}\epsilon^{acd}\;\epsilon^{bef} \;
\underline{F}_{cd}^J\;\underline{F}_{ef}^K]\delta_{JK} 
\nonumber\\
C^F &=& i\frac{E^a_j}{|\det(E)|^{1/2}}\; [(\eta^T)^\ast\;\tau_j\; (\nabla_a \eta) - c.c.]
+m(\eta,\eta^\ast)
\nonumber\\
C^S &=& \frac{1}{2}\;\frac{1}{|\det(E)|^{1/2}} \; \{[(\pi^\iota)^\ast;\pi^\kappa+
E^a_j E^b_k\delta^{jk} \; (\nabla_a \phi)_\iota^\ast\; (\nabla_b \phi)_\kappa]\delta^{\iota\kappa}
+|\det(E)|\; V((\phi^T)^\ast \phi)\}
\nonumber\\
C^Y &=& [(\phi_\iota)^\ast \; (\nu^A)^\ast\; \delta_{AB} \; \delta_{\iota\kappa} \; \eta^B_\kappa
+c.c],\;\;
C^D = h^D(X,Y,E)
\ea
Here $F, \;\underline{F}$ are the curvature 2-forms of $A,\;\underline{A}$ respectively,
$K_a^j:=A_a^j-\Gamma_a^j$ is a derived field depending on both $A,E$ closely related 
to the extrinsic curvature, $\beta$ is numerical real valued and non vanishing 
constant depending on the  Immirzi parameter \cite{13} (which we set here to unity for simplicity
of exposition), $J_j=(\eta^T)\ast \tau_j \eta,\; 
\underline{J}_J=(\eta^T)\ast \underline{\tau}_J\eta$ are purely fermionic currents bilinear in 
the fermion fields, $\epsilon_{jkl}=(T_k)_{jl}, f_{JKL}=(\underline{T}_K)_{JL}$ are the 
respective structure constants, $m(\eta,\eta^\ast)$ denotes a general, gauge invariant bi-linear 
mass term of Dirac and/or Majorana type and $V$ a gauge invariant polynomial 
potential (including mass terms). The suffix at the various contributions to $C$ 
refers to the various geometry and matter species: Euclidian and Lorentzian geometry
term, cosmological term, Yang-Mills term, fermionic term, scalar term, Yukawa term
and dark matter or dust term. To obtain the desired phenomenology one can add more 
field species with corresponding mixing matrices and gauge groups. The common feature 
of the $C^D$ contribution is that, in the models considered so far, it depends only on 
the dark matter or dust fields and the spatial geometry encoded by $E$. 
We also have set various coupling constants to unity for simplicity of presentation.
see \cite{13a-5} for more details.\\
\\
The main reason for displaying these various contributions to the constraints explicitly 
is to draw attention to the following: 
\begin{itemize}
\item[1.] {\it Pure (Euclidian) vacuum GR is unphysical}\\
Except for the dark matter term and the inflaton terms all pieces in the above list are 
experimentally confirmed, the signature of the universe is Lorentzian and not Euclidian.
Therefore a theory of quantum geometry must ensure that all contributions to
 $G,\underline{G},D,C$
are taken into account and not just bits and pieces of it.
\item[2.] {\it Non-polynomiality}\\
The non-polynomial character of the constraints is present only in the Hamiltonian 
constraint and is only due to the non-polynomial appearance of the field $E^a_j$ 
which is a triad of density weight one. One may be tempted to multiply $C$ by  
a sufficiently high power $r$ of the density weight unity quantity 
\be \label{3.2b}
Q:=|\det(E)|^{1/2}
\ee
in order to turn it into a 
polynomial. In the classical theory this is possible because here $Q>0$ is 
an implicit assumption and thus the constraints $C,\; Q^r \; C$ define the same
constraint surface. 
This is due to the fact that the relation to the intrinsic 
metric is given by 
\be \label{3.2a}
q^{ab}\det(q):=E^a_j\; E^b_j\;\delta^{jk} 
\ee    
In the quantum theory this equivalence is no longer manifest if $Q>0$ is 
not granted also there. As we will see below, apart from that, quantum theory by itself 
dictates that one keeps the {\it density unity} constraint $C$ as it is in its 
{\it non-polynomial} form.
\item[3.] {\it Dictated density weights of the elementary fields}\\
The explicit form of $D_a$ shows that the various fields have the following 
density weights that naturally come out of the Legendre transform:
$A, \underline{A}, \phi, X$ have density weight zero,
$E, \underline{E}, \pi, Y$ have density weight one,
$\eta, \nu$ have density weight $1/2$.
\item[4.] {\it Universal coupling of $E$}\\
We see explicitly that that the field $E$ {\it appears in every single term} of 
the Hamiltonian constraint, it couples to $A$ and matter 
in various different forms:
\be \label{3.2c}
\frac{\epsilon^{jkl} E^a_k E^b_l}{|\det(E)|^{1/2}},\; \delta_{jk} E_a^j E_b^k,\; 
\frac{E^a_j}{|\det(E)|^{1/2}},\;
\frac{1}{|\det(E)|^{1/2}},\; |\det(E)|^{1/2}, 
\frac{E^a_j E^b_k \delta_{jk}}{|\det(E)|^{1/2}},\; 
\ee
Here the field 
$E_a^j$ is the {\it inverse} of $E^a_j$, i.e. $E^a_j E_a^k=\delta_j^k$. The list 
(\ref{3.3}) is incomplete because we have to remember that $K_a^j=A_a^j+\Gamma_a^j$, 
and $\Gamma_a^j$, which also appears in  the fermionic $\nabla \eta$, is of the 
form $\frac{E \cdot E \cdot\partial E}{\det(E)}$. This again shows 
that it is a {\it necessary assumption} in the classical theory that $E$ be 
{\it non-degenerate}, in particular $Q:=|\det(E)|^{1/2}>0$ everywhere and every time.
\end{itemize}

\subsection{Poisson algebra of constraints}
\label{s3.3}

As follows from the abstract argument in \cite{5} and as one can also verify by tedious 
calculation using the canonical brackets between the list of fields in (\ref{3.1}), 
the constraints of the previous subsection obey the subsequent universal (i.e.
independent of the Lagrangian) Poisson 
algebra
\ba \label{3.4}
\{(G,\underline{G})[L,\underline{L}],\;(G,\underline{G})[L',\underline{L}'],\;
&=& -(G,\underline{G})[[L,L'],[\underline{L},\underline{L}']]
\nonumber\\
\{(G,\underline{G})[L,\underline{L}],\;D[u]\} &=&
-\{(G,\underline{G})[u[L],u[\underline{L}]]
\nonumber\\
\{(G,\underline{G})[f,\underline{f}],\;C[f]\} &=& 0
\nonumber\\
\{D[u],\; D[u']\} &=& -D[[u,u']]
\nonumber\\
\{D[u],\; C[f]\} &=& -C[u[f]]
\nonumber\\
\{C[f],\; C[f']\} &=& - D[q^{-1}[f\; df'-f'\; df]]
\ea
where 
\be \label{3.5}
(G,\underline{G})[L,\underline{L}]=\int_\sigma\; d^3\; [L^j \; G_j+\underline{L}^J
\; \underline{G}_J],\;\;
D[u]=\int_\sigma\; d^3\;u^a\; D_a,\;\;
C[f]=\int_\sigma\; d^3\;f\; C
\ee
are the smeared constraints. Here $[L,L']$ denotes the commutator in the Lie algebra
of SU(2) and similar for the group $G$, $[u,u']$ is the commutator in the Lie algebra
of vector fields and $u[f]$ is the action of the derivation $u$ on the scalar $f$.

The subalgebra spanned by the $D[u], C[f]$ alone is known as hypersurface deformation
algebra and it is not a Lie algebra due to the presence of the field $q^{-1}$ displayed
in (\ref{3.2a}) in the structure functions.

\subsection{Kinematical Hilbert space representations}
\label{s3.4}

Although in \cite{13} an elegant argument based on diffeomorphism co-variance is presented
which establishes that the kinematical Hilbert space representation is essentially unique,
we give here a shorter argument based on the dynamics of the theory displayed above.

In QFT on Minkowski space one usually splits the polynomial Hamiltonian into quadratic and 
higher order part and uses the quadratic part to select a (Fock) vacuum. 
Then the quadratic part is densely defined on the resulting free or ``kinematical'' 
Fock space. The chosen Fock vacuum also defines what one means by normal ordering and in that sense also severely 
affects the higher order part. In our non-perturbative and non-polynomial setting 
a natural split into quadratic and higher order is not available. However,
the fact that the field $E$ couples to every single term in the Hamiltonian constraint
in algebraically different forms 
strongly motivates to select a vacuum $\Omega$ annihilated by $E$ in order that each of those 
terms be individually densely defined. This also requires inverse powers of $Q$ 
(\ref{3.2b}) to annihilate the vacuum which is possible as we will see.

We now show that the innocent looking condition $E^a_j\; \Omega=0$ has far reaching 
consequences \cite{13g}. As it is customary in QFT, instead of considering the 
abstract Heisenberg algebra generated by the relations (exemplified for geometry) 
\be \label{3.6}
[E[f],E[g]]=[F[A],F'[A]]=0,\;[E[f],F[A]]=i{\bf 1},\;
E[f]^\ast-E[f]=F[A]^\ast-F[A]=0
\ee
with $F[A]:=\int_\sigma\; d^3x\; F^a_j\; A_a^j,\;
E[f]:=\int_\sigma\; d^3x\; f_a^j\; E^a_j$
where $f,F$ are real valued test functions, we consider the Weyl algebra 
generated by Weyl elements $U[F]=\exp(i\; F[A]),\;V[f]=\exp(i\; E[f])$ and induced 
relations
\ba \label{3.7}
&& U[F]\;U[F']=U[F+F'],\;U[F]^\ast=U[-F],\; 
V[f]\;V[f']=V[f+f'],\;V[f]^\ast=V[-f]
\nonumber\\
&& V[f]\; U(F)\; V[-f]=U[F]\; e^{-i\;F[f]}
\ea
In any representation, the Weyl elements are unitary i.e. bounded operators so that 
no domain questions arise as compared to the generators of the Heisenberg algebra.
It now follows from our assumption $V[f]\Omega=\Omega$ for all $f$ that
\be \label{3.8}
<\Omega,\; U[F]\; \Omega>= 
<V[-f]\Omega,\; U[F]\; V[-f]\Omega>=e^{-i\;F[f]} \; <\Omega,\; U[F]\; \Omega>
\ee
for all $f,F$. It follows that the representation is necessarily of {\it 
Narnhofer-Thirring type} \cite{30}
\be \label{3.9}
<\Omega, U[F]\Omega>=\delta_{F,0}
\ee
where $\delta$ really means the Kronecker function. This means that
the unitary operators $U[F]$ are not strongly or weakly continuous and that 
the states $U[F]\Omega$ are orthonormal implying that the resulting Hilbert space 
spanned by the $U[F]\Omega$ is not separable. 

The LQG representation uses more complicated functions than the $U[F]$ based on 
holonomies of $A$ along 1-dimensional curves as it is motivated by Gauss gauge covariance but the main 
argument above still applies. We will see in a moment that besides gauge covariance 
again purely dynamical arguments dictate that $F$ should smear $A$ in one dimension and 
that $f$ should smear $E$ in 2 dimensions.

As far as the matter representations are concerned, motivated by background 
independence which excludes Fock typ representations, one can similarly pick representations
defined by the requirements $\underline{E}\Omega=\pi\Omega=\eta\Omega=\nu\Omega=Y\Omega=0$ 
\cite{13a-6}.
For the YM, scalar and DM  sector the corresponding representations are again discontinuous,
the smearing dimensions in the YM sector are the same as in the geometry sector 
while for the scalar sector $\phi,\pi$ and $X,Y$ are smeared in 0 and 3 dimensions respectively.
We can therefore pick a Hilbert space representation for the YM sector analogous to the 
geometry sector and for the scalar sector an analogous representation based on ``point
holonomies''. 
For the fermionic sector we obtain the ususal continuous Fock representation familiar 
from the standard model quantisation and the natural smearing dimension is $3/2$.
In this way, scalar and fermionic matter is excited at 
points while geometry and YM degrees of freedom is excited along paths. A combined 
generalised SNWF for all degrees of freedom is thus labelled by a graph with 
irreducible representations of both SU(2) and G labelling the edges corresponding 
to the geometry and YM field excitations, irreducible representations for SU(2) at the 
vertices for the fermions and irreducible representations at the vertices for the 
fermions and Higgs scalars as well as gauge invariant intertwiners at the vertices, see 
\cite{13a-6} for details and subsection \ref{s3.1}.

We will see below that also those representations and smearing dimensions are naturally
suggested by the dynamics.

\subsection{Smearing dimensions and density weights}
\label{s3.5}

The main mathematical problem of QFT is to define products of operator valued distributions.
In the density weight one form, the problem in quantum gravity is even worse, as 
one also has to deal with inverse powers of fields. Yet, it is precisely that 
non-polynomial nature of the constraints which offers the possibility to arrive at a 
non-perturbative quantisation. 

As it is customary in QFT we try to tame the products of distributions by introducing 
a short distance cut-off $\epsilon$. This is conveniently obtained using a e.g. simplicial
decomposition $\cal T$ of $\sigma$ into tetrahedra $\Delta$ of coordinate volume of order 
$\epsilon^3$. The classical Hamiltonian constraints can now be written as 
\be \label{3.10}
C_\epsilon[f]=\sum_{\Delta\in {\cal T}}\; C_\epsilon^\Delta[f],\;\;
C_\epsilon^\Delta[f]=\int_\Delta\; d^3x\; f\; C
\ee
which is still exact.
If $p_\Delta$ denotes the barycentre of $\Delta$ in the chosen system of coordinates 
then we have (up to numerical constants)
\be \label{3.11}
C_\epsilon^\Delta[f]\approx \; f(p_\Delta)\; [C(p_\Delta) \epsilon^3]
\ee
The challenge is now to approximate the continuum fields that appear in 
$C(p_\Delta)\; \epsilon^3$ 
by smeared versions that have well-defined operator substitutes on the kinematical Hilbert 
space such that the following holds:\\
1. The resulting classical expression $\hat{C}_\epsilon^\Delta[f]$ differs from 
$C(p_\Delta)\; \epsilon^3$ by a term of order $\epsilon^4$\\
2. When the operator substitution in  $\hat{C}_\epsilon^\Delta[f]$ is carried out
the operator $\hat{C}_\epsilon[f]=\sum_\Delta \hat{C}_\epsilon^\Delta[f]$ is densely 
defined on the kinematical Hilbert space with dense, invariant domain independent of 
$f$\\
3. The limit $\epsilon\to 0$ of $\hat{C}_\epsilon[f] \psi,\; \psi\in D$
exists and is not trivial for every $\psi\in {\cal D}$.\\
Requirement 1. just makes sure that the classical $\hat{C}_\epsilon^\Delta[f]$ is an 
admissable disretisation. Requirement 2. is the main point of this point splitting 
regularisation, namely to give the regulated expression a mathematical meaning. Insisting
that the domain is invariant and independent of $f$ is to enable computing commutators.
Requirement 3 encodes the removal of the regulator defining the regulator independent 
operator densely on $\cal D$ and such that it is not the trivial, zero operator.

These requirements together with the concrete form of the Hamiltonian 
constraint now uniquely fix the (integer) 
smearing dimensions of the elementary fields as well
as that changing $C$ to $Q^r C$ for some real number $r\not=0$ is not allowed. 
The key is to realise that the factor $\epsilon^3$ universally multiplies every single 
term of $C(p_\Delta)$, i.e. when rewriting the ``naked'' fields at the point 
$p_\Delta$ in terms of their smeared versions the factor $\epsilon^3$ must 
be precisely absorbed by every single matter and geometry contribution.

We begin with the cosmological term with density weight $w$. In order that 
$Q^w(p_\Delta) \epsilon^3$ satisfies requirements 2, 3 the expression 
$\epsilon^{2/w} E^a_j(p_\Delta)$ must be substitutable by a well defined smeared 
field operator, hence
the smearing dimension of $E$ is $2/w$.

Next we consider the spatial curvature term which is implicitly contained in $C^G_L$ and 
is given by $\epsilon^3\;Q^{-1}(\Gamma_a\times \Gamma_b)\cdot (E^a\times E^b)$
plus a similar term with $\Gamma^2$ replaced by $\partial \Gamma$. From 
the definition of the spin connection  
\be \label{3.12}
(\nabla_a E)^b_j=
(\nabla_a E_j)^b+\epsilon_{jkl}\;\Gamma_a^k E^b_l=0
\ee 
where the first term just acts on the tensor structure of $E$, we see that $\Gamma_a^j$
is schematically of the form $Q^{-2}\; E\; E \; \partial E$. Thus the spatial curvature
term is schematically of the form $\epsilon^3\; Q^{-5} \; E^6\; (\partial E)^2
=\frac{[\epsilon^3 \partial E]^2}{\epsilon^3 Q}$. As $Q\epsilon^3$ is already well 
defined by the the paragraph before (\ref{3.12}), $\epsilon^3 \partial E$ must be well defined. As $\partial$ increases
the required smearing dimension by one unit (integrate by parts to see this), 
$\epsilon^2 E$ must be well defined, i.e. the smearing dimension of $E$ is two. 
It follows unambiguously that $w=1$ and that the above discussed similar term
is also well defined.

Next we consider the Euclidian geometry term $C^G_E$ at density weight 
$w=1$ which in particular contains a term of the 
schematic form $\epsilon^3\;(A\; A\; E\; E \;Q^{-1})(p_\Delta)$.
The factor $\epsilon^3$ must then be distributed as 
$(A\epsilon)^2\; \; (E\epsilon^2)^2\; (Q\epsilon^3)^{-1}$, i.e. the smearing dimension 
of $A$ must be unity.

Continuing this reasoning in the various matter contributions to $C$ we find unambiguously
that $(\underline{A},\underline{E})$ have smearing dimension 1,2 respectively, 
that $(\phi,\pi)$ and $(X,Y)$ have smearing dimension 0,3 respectively and that
$(\eta,\nu)$ have smearing dimension $\frac{3}{2}$ respectively. 
The fact that all of this fits together although the algebraic form of the various
terms in $C$ is quite different is quite remarkable and is due to the diffeomorphism
covariance of the theory.

\subsection{Inverse powers of $E$ and quantisation ambiguities}
\label{s3.6}

From the form of $C$ it is clear that inverse powers of $E$ only appear in the form
$[\epsilon^3 Q]^{-n}$. In order to make these well defined one should make sure that 
1. the volume of regions $R$ given classically by $V(R)=\int_R\; d^3x\; Q$ becomes a
well defined operator and 2. that $V(R)^{-n}$ can be given a meaning. The first 
task is precisely accomplished in the chosen representation \cite{33a,33b,33c} as reviewed in 
\cite{13}. 

The second task can be accomplished in many different ways, see \cite{13a-5}
for a complete classification. The first possibility rests on the Poisson bracket 
identity 
\be \label{3.13}
\{V(R),A_a^j(x)\}=[Q\; E_a^j](x)\;\;\Rightarrow\;\;\det(\{V(R),A(x)\})=Q
\ee
for any $x\in R$. Thus for any region of coordinate size $\epsilon^3$
\be \label{3.14}
V(R)^{-n}=V(R)^{-[n+m]} [\det(\{V(R),A_e(p)\}]^m=\frac{1}{r^3}\;
[\{\det(V(R)^r,A(e)\}]^m,\; 3(r-1)m=-(m+n)  
\ee
where $p\in R$, $e=(e_1, e_2, e_3)$ are the axes of a local coordinate system in
$R$ and $A^j_{e_I}(p)=\int_{e_I\cap R}\; A^j,\;I=1,2,3$. The equality (\ref{3.14}) is 
modulo terms of order higher than $\epsilon^3$. For any $m$ such that $(m+n)/(3m)<1$ 
i.e. $m>n/2$ we have $0<r<1$ thus the r.h.s. of (\ref{3.14}) depends on 
{\it positive} powers of $V(R)$. Since also to order $\epsilon$ we have 
${\rm Tr}(h_{e_I}(A)\tau_j)=A^j_{e_I}(p)$ where $h_e$ denotes the holonomy of $A$ along $e$,
we can turn (\ref{3.14}) into a well defined operator by replacing the classical 
objects $V(R),h_e$ by their operator equivalents and the Poisson brackets 
into commutators divided by $i$ (if we set $\hbar=1$) by using the spectral 
theorem on the self-adjoint positive operator $V(R)$.          

The second possibility is the Tychonov regularisation \cite{34}
\be \label{3.15}
V(R)^{-n}:=\lim_{\delta\to 0}\; [\frac{V(R)}{\delta^2+V(R)^2}]^n
\ee
Common to both possibilities is that with this definition, and if one {\it orders the
dependence on the volume operator to the outmost right}, every contribution to $C$
only acts in the vicinity of a vertex of a SNWF $T_\gamma$ over $\gamma$ 
because $V(R)\;T_\gamma=\sum_{v\in V(\gamma)\cap R}\; V_v\; T_\gamma$ 
where $V(\gamma)$ is the 
set of at least tri-valent (with respect to the geometry degrees of freedom)
vertices of $\gamma$ and $V_v$ a densely defined 
local operator. Thus only those tetrahedra $\Delta$
of the simplicial decomposition contribute which contain a tri-valent vertex of 
$\gamma$. It is then natural to adapt the simplicial decomposition to $\gamma$ so 
that $v=p_\Delta$ is that vertex if $\Delta$ contains one (for $\cal T$ sufficiently fine,
each $\Delta$ contains at most one)
and the edges $e_I$ appear as segments of edges
of $\gamma$ adjacent to $v$. See \cite{13a-1a,13a-1b} for more details.
That the action of $C$ will eventually restrict to the vertices of $\gamma$ will 
make sure that the resulting operator is densely defined in the SNWF basis.
Again it is quite remarkable that this is possible at all, given the tremendous 
degree of non-polynomiality of the constraint $C$.

We note that inverse volume powers give rise to a substantial amount of regularisation
ambiguities in addition to ordering ambiguities: First because we may pick any $m>n/2$
or may choose to write $n=n_1+..+n_r$ and treat each inverse power $n_k,\;k=1,..,r$ 
individually and because instead of picking the spin $1/2$ representation
in the approximation  ${\rm Tr}(h_{e_I}(A)\tau_j)=A^j(e_I)$ we may pick any spin $k$
rep ${\rm Tr}(\pi_k(h_{e_I}(A))\tau^{(k)}_j)=d_k\; A^j(e_I)$ where $d_k=2k+1$ is its
dimension and $\tau^{(k)}_j$ the Lie algebra basis element in that representation
\cite{35}. These ambiguities arise because the constraints are not polynomial so that 
the ``principle of simplicity'' appears less natural than in regularisations of usual
QFT's.

\subsection{Complete regulated operator}
\label{s3.7}

We sketch the regularisation of the individual terms of $C$ referring to \cite{13a-5} 
for details.

The piece $C^G_E$ requires the quantisation of $F$. Since $A$ or $F$ do not exist 
as operator valued distributions in this representation (due to the discontinuity
of Weyl elememts mension in section \ref{s3.4}) one replaces $\epsilon^2 F$ by the 
holonomy along a loop starting in a vertex $v$ of $\gamma$ along the segments of two 
adjacent edges of $\gamma$ whose end points are connected by a new edge. This is why
the action is called graph changing. This involves a sum over pairs of such edges.
In \cite{36,37} this ``loop attachment'' is chosen not along existing edges but 
``close but disjoint'' from those. This has the advantage of a simpler solution structure
but does not lead to propagation \cite{28} in the space of solutions.

The piece $C^G_L$ requires the quantisation of $\epsilon K_a^j=\epsilon(A_a^j-\Gamma_a^j)$. 
This can be done using the classical Poisson bracket identity 
\be \label{3.16}
\epsilon K_a^j=\{\{C^G_E[1],V(\sigma)\},\epsilon\; A_a^j\}
\ee
and replace classical objects by quantum counter parts and Poisson brackets by commutators
divided by $i$ (also $\epsilon A$ is substituted by a holonomy) using that $C^G_E$ is 
already defined. Another possibility is based on quantising the spin connection 
$\Gamma$ itself \cite{38}.

The piece $C^C$ is essentially $V(\sigma)$ which is already quantised.

The piece $C^{YM}$ replaces $\epsilon^2 \underline{E}$ by a YM flux operator and 
$\epsilon^2 \underline{F}$ by YM loop holonomies in complete analogy to $(E,A)$ while the $E$ 
dependent terms are treated according to section \ref{s3.6}.

The pieces $C^F$ replaces $\epsilon^{3/2} (\eta,\nu)$ by Fock operators located 
at vertices and otherwise treats the ingredients $E,A,K$ as before.

The piece $C^S$ replaces $\epsilon^3 \pi(v)$ by i times 
an ordinary derivative w.r.t. $\phi(v)$ while 
$2\epsilon [\partial_a \phi](v)$ is replaced by 
$[e^{i\phi(v+\epsilon \delta_a)}-e^{i\phi(v-\epsilon \delta_a)}]\; e^{-i\phi(v)}$
in case that $\phi$ transforms in the trivial or adjoint representation (if it
transforms in the defining representation as in the SM, one should first 
carry out the Higgs mechanism  classically and reduce the treatment to the trivial 
representation)
where $\delta_a$ denotes a translation in local a-direction and $\phi(v)-\phi(\nu_0)$ itself is replaced 
by an $\epsilon$ resolution Riemann sum approximation of 
$i^{-1}\int_{c_v} dW(x)\; W(x),\; W(x)=\exp(i\phi(x))$ where $c_v$ is a path from some 
reference point $v_0$, at which $\phi$ decays to zero or is otherwise fixed by the classical boundary
conditions, to $v$. Alternatively one may instead work with the more general 
functions $W_\mu(x)=e^{i\mu \phi(x)}$ and replace $\phi(x)$ by 
$[W_\mu(x)-W_{-\mu}(x)]/(2i\mu)$ for fixed $\mu$ \cite{39}. The then 
introduced additional dependence of 
the operator on $\mu$ is much debated in the Loop Quantum Cosmology (LQC) literature
\cite{40a,40b,40c,41a,41b} which focuses on the cosmological sector of the theory 
and uses a lot of the technology introduced above for the full theory.

The piece $C^Y$ uses all of the above.\\
\\
An essential feature of the matter contributions is the following: {\it matter can 
be present only where geometry is excited}. This means that the action of the 
matter contributions is also restricted to those vertices of the generalised SNWF 
which carries gravitational volume. This physically quite plausible fact which comes 
out naturally from the formalism plays an important role for the closure of the
algebra, see below.   
 
It is transparent that the complete quantisation of the regulated operator has introduced 
additional quantisation ambiguities (e.g. the details of the loop attachment) and 
the question is how much of that survives upon removal of the regulator $\epsilon$.

\subsection{Gauss constraint and spatial diffeomorphism constraint}
\label{s3.9}

Before we discuss the regulator removal for the Hamiltonian constraint, we first 
construct the quantum operators corresponding to the remaining constraints as well
as their solutions.

As shown in \cite{42a,42b,42c,42d} whenever 1. the constraints are linear in momentum and 2. 
those momenta annihilate the vacuum, which is the 
case for these constraints, one can obtain unitary operators 
$U(L,\underline{L},u)$ corresponding to their exponentiation 
$\exp(i[G[L]+\underline{G}[\underline{L}]+D[u]])$ defined densely 
by 
\be \label{3.17}
U(L,\underline{L},u)\; w[F^G,\underline{F}^{YM},F^F,F^S]\Omega
=w[(e^X_{L,\underline{L},u}\cdot K)((0,F),(0,\underline{F}),(0,F^F),(0,F^S)]\Omega
\ee
where $X_{L,\underline{L},u}$ is the Hamiltonian vector field of 
$G[L]+\underline{G}[\underline{L}]+D[u]$, the Weyl element is given as before by 
i times the path ordered exponential of
$F[A]+\underline{F}[\underline{A}]+F^F[(\eta,\nu)]+F^S[\phi]$ 
where $F^\ast$ denote the smearing functions (real or Lie algebra valued for bosons
see section \ref{s3.4}, Grassmann 
valued for fermions, see also \cite{13a-5} for more details) of the respective sectors 
with the smearing dimensions derived above and $K$ is the momentum coordinate function
on the classical phase space. Therefore (\ref{3.17}) is natural, free of any ambiguities and those constraints 
close among themselves without anomalies as they should. 

Note that the generators of 1-parameter subgroups
exist for the Gauss constraints but not for the spatial diffeomorphism constraints
due to the discontinuity of the representation. Thus while the solution of the 
Gauss constraints simply restrict the kinematical Hilbert space to its Gauss invariant 
subspace (Gauss invariant SNWF) which just requires harmonic analysis on 
SU(2) and G \cite{43}, the solutions of the spatial diffeomorphism constraints 
are distributions ${\cal D}^\ast_{{\rm diff}}$ that result from averaging over all diffeomorphisms, see e.g. 
\cite{44} where also possible Hilbert space structures on those spaces of distributions 
are discussed.

\subsection{Regulator removal from the Hamiltonian constraint and operator topologies}
\label{s3.10}

The removal of the $\epsilon$ dependence requires the specification of an operator 
topology, i.e. a notion of convergence. The standard weak or strong operator topologies 
cannot be used because e.g. holonomy operators are not weakly or strongly continuous
with respect to the path along which they are defined. Apart from the spaces $\cal D$ and 
$\cal H$ the only other natural spaces available at this point are the space ${\cal D}^\ast$
of algebraic distributions over $\cal D$ (linear functionals without continuity notion) and 
its subspace ${\cal D}^\ast_{{\rm diff}}$ of spatially diffeomorphism invariant 
elements discussed in the previous subsection. This suggests to consider a topology that 
is reminiscent of the so-called weak$^\ast$ operator topology \cite{45}: An open 
neighbourhood base for this topology is given by 
\be \label{3.18}
N_\delta(l_1,..,l_m;\; T_1,..,T_n;\;O)=\{O';\; |l_r[(O'-O)T_s]|<\delta;\; 1\le r\le m;
1\le s\le n\}
\ee
where $O,O'$ are in the set of operators with $\cal D$ as dense invariant domain 
and $l_r\in {\cal D}^\ast_{{\rm diff}},\; T_s\in {\cal D}$. Early steps towards 
this topology were stated in \cite{46}, see \cite{13a-1a,13a-1b,13a-2} for more details.

As the loop attachment is performed in a spatially diffeomorphism co-variant way, 
one finds that for any smearing function $f,\; l\in {\cal D}^\ast_{{\rm diff}},\;
T\in {\cal D}$ that 
\be \label{3.19}
l[(C_\epsilon[f]-C_{\epsilon_0}[f])\;T]=0
\ee
for any $\epsilon,\epsilon_0$ where it is understood that for $\epsilon>0$ the prescription 
of loop attachments and similar finite size ambiguities is such it overlaps with the given 
graph except for an additional ``arc'' between two segments of edges adjacent to a given vertex 
which is attached transversally to their endpoints and
that intersects the graph nowhere else and whose braiding is the same for all $\epsilon$.
Using the axiom of choice we can now pick once and for all some $\epsilon_0$ and 
{\it define the regulator free operator} by $C[f]:=C_{\epsilon_0}[f]$.

\subsection{Commutator algebra, closure and anomalies}
\label{s3.11}

We can now check whether the hypersurface deformation algebra $\mathfrak{h}$ is represented without 
anomalies. Actually, it is not possible to do this for $\mathfrak{h}$ itself, because the 
operator $D[u]$ does not exist in the chosen representation. We will therefore verify a 
classically equivalent algebra generated by $\exp(X_u),X_f$ where $X_u, X_f$ 
are the Hamiltonian vector fields of $D[u],C[f]$ respectively. We have by the general 
relation between Hamiltonian vector fields and Poisson brackets from (\ref{3.4})
\be \label{3.20}
e^{X_u}\; e^{X_v}\; e^{-X_u}=\exp(X_{e^{[u,.]}\cdot\;v}),\;
e^{X_u}\; X_f\; e^{-X_u}=X_{e^u\;\cdot\;f},\;
[X_f, X_g]=X_{u=-q^{-1}(f\; dg-g\;df)}
\ee
The last relation {\it cannot} be written in the form of $e^{X_w}$ for some, possibly 
phase space depeendent vector field  $w$, fo that one would also have to eponentiate the Hamiltonian 
constraint which is in fact possible for the U(1)$^3$ truncation of Euclidian vacuum
GR with cosmological constant \cite{42a,42b,42c,42d}. Fortunately, in contrast to $D[u]$ itself 
the quantity $D[q^{-1}(f\; dg-g\;df)$ {\it can be quantised} in the chosen representation
\cite{13a-3}, hence we leave (\ref{3.20}) as it stands.

In the quantum theory we find
\ba \label{3.21}
&& U(u)\; U(v)\; U(u)^{-1}=U(e^{[u,.]}\;v),\;
U(u)\; C[f]\; U(u)^{-1}\; T_\gamma=C[e^{u\cdot f}]\;T_\gamma+
[\tilde{U}(\varphi_{u,f,\gamma})-1]\; T_\gamma,\;
\nonumber\\
&& [C[f],C[g]\; T_\gamma=\sum_{v,v'\in V(\gamma)} [f(v) g(v')-f(v')\; g(v)]\;
[\tilde{U}(\varphi_{\gamma, v, v'})-1]\; 
C_{\gamma_{v},v'}\; C_{\gamma,v}\; T_\gamma 
\ea
where $C_{\gamma,v}$ is the contribution from the vertex $v$ when acting on $\gamma$
and $\gamma_v$ is the graph resulting from this action (we are over simplifying here,
see \cite{13a-2} for the details).
Here $\tilde{U}$ is the extension of $U$ from diffeomorphisms generated by vector fields 
$u$ to general diffeomorphisms $\varphi$ and the diffeomorphisms displayed depend on the 
structures indicated. Note that the double sum in the second line involves only the vertices 
of $\gamma$ and not the new vertices resulting from the first action which is due 
to the fact that these vertices are co-planar and annihilated by the employed version of 
the volume operator. This is also the reason 
for why this works for all matter couplings, the cosmological constant and both 
vacuum GR contributions which are all contributing only where geometry is excited.

Thus the diffeo -- diffeo commutator is anomaly free, the 
diffeo -- Hamiltonian commutator is anomaly free modulo a term
proportional to the exponentiated diffeomorphism constraint $\tilde{U}(\varphi)-1$
and the Hamiltonian - Hamiltonian 
commutator is plain anomalous: it is non-vanishing, and while proportional to linear combinations 
of the exponentiated  
diffeo constraint correctly ordered to the outmost left and in that sense 
closes, i.e. does not lead to new constraints, the ``operators of proportionality''
or quantum structure functions are incorrect: as announced, a possible quantisation 
of $D[q^{-1}(f\;dg-g\;df)]$ is given by (its regulator free version is obtained in the 
same topology as for the Hamiltonian constraint and is non-vanishing for
discontinuous $f,g$) \cite{13a-3}
\be \label{3.22}
D[q^{-1}(f\;dg-g\;df)]\;T_\gamma=-\sum_{v\in V(\gamma)}\sum_{e\cap e'=v}
[f(v)\;g_e(v)-f_e(v_)\;g(v)]\;[\tilde{U}(\varphi_{\gamma,v,e'})-1]\; Q^{e,e'} T_\gamma
\ee
where $\varphi_{\gamma,v,e'}$ is a diffeomorphism with support in the vicinity of $v$
generated by a vector field which coincides on $e'$ with the tangent vector field of 
$e'$, $Q^{e,e'}$ is geometrical operator of the form discussed in subsection
\ref{s3.6} which is a quantisation of $q^{ab}$ and $f_e(v):=\lim_{t\to 0} f(e(t))$ 
is the path dependent limit of $f$ at $v$ with $e(0)=v$. Formula (\ref{3.22}) displays 
the expected ``quantum structure functions which differ from those in (\ref{3.21}).
Note that in the classical theory one works with continuous, typically 
even smooth, functions $f,g$ so that the right hand side of (\ref{3.22}) would 
be zero. The reason for this discrepancy is that in the quantisation of the constraints a central ingredient of the classical theory, namely the non-degeneracy
of the spatial metric is violated. If one would reinstall non-degeneracy, the 
discrepancy is likely to disappear. This is furtther discussed in section \ref{s3.17}.

One reason 
for the failure of the last relation in (\ref{3.21}) to produce (\ref{3.22}) is due to 
the fact that the AL volume operator chosen in the quantisation
vanishes on the co-planar tri-valent vertices 
produced by the first action of the Hamiltonian constraint. On the other hand, 
if it would not (e.g. by using the RS volume) the commutator would not even be 
a linear combination of exponentiated diffeo constraints. At least in the present 
form the Hamiltonian constraint does not enforce solutions to the quantum constraints 
in addition to the exponentiated diffeo and Hamiltonian constraints and in that sense 
the type of the anomaly is less disastrous than one which leads to downsizing the number of 
physical degrees of freedom more than in the classical theory. However, 
improvement of the action of the Hamiltonian constraint on $\cal D$ must be such that 
a second action does not vanish at the vertices produced by the first action which is 
not ruled out to be possible but has not been done yet. See \cite{48a} for a toy model 
where this step could be completed.

To conclude, we distinguish between the notions of {\it closure} and {\it non-anomalous}:
Closure means that a quantum commutator algebra of constraints is a linear combination 
of those constraints ordered to the left with {\it some} structure operators.  
Non-anomalous means that those structure operators qualify as the quantisation of the 
corresponding classical structure functions. The constructed algebra thus closes 
but is anomalous. This is better than non-closure but still a non-anomalous representation
is desired.   

\subsection{On-shell closure, off-shell closure and habitats}
\label{s3.11a}
  
By definition, an algebra closes off-shell on some {\it invariant} space if that space contains 
elements not annihilated by all algebra elements. It closes partly (fully) on-shell if the space 
consists of elements which are annihilated by a subalgebra (the full algebra - in that case 
the action of the algebra on the space is trivial).

The algebra of the $\tilde{U}(\varphi)-1,C[f]$ that we have defined in the previous 
subsection acts on the dense invariant domain $\cal D$ which is not annihilated by 
either the $C[f]$ or the $\tilde{U}(\varphi)-1$. It is therefore an {\it off-shell, non-Abelian}
representation which however is anomalous. This fact is often confused in the 
literature, see e.g. \cite{47}: The space ${\cal D}^\ast_{{\rm diff}}$ is just used to define 
a topology wrt which one can remove the regulator 
$\epsilon$ in $C_\epsilon[f]$ i.e. to reach the limit $C[f]$. The operator is defined 
on $\cal D$ and not on ${\cal D}^\ast_{{\rm diff}}$, it cannot possibly be because 
$C[f]$ is not diffeomorphism invariant so ${\cal D}^\ast_{{\rm diff}}$ cannot possibly 
be a an invariant domain. See \cite{48} for more details.

Still one can try to define $C[f]$ on a different space as first suggested in \cite{13e,13e2}:
It is a subspace of ${\cal D}^\ast$ containing ${\cal D}^\ast_{{\rm diff}}$ which however
is genuinely larger than it. Therefore the action of the algebra on that space 
${\cal D}^\ast_{{\rm vs}}$ of so-called vertex smooth distributions or ``habitat'' 
will be off-shell.
The space has a basis whose elements $l$ consist of linear combinations of SNWF with the same    
spin and intertwiner label but where we sum over the diffeomorphism class of the 
graph label with graph dependent 
coefficients. If those coefficients depend continuously on the graph then 
the definition $(C'[f]\;l)[T]:=\lim_\epsilon l(C_\epsilon[f]\; T)$ is well defined and 
$C'[f]$ leaves ${\cal D}^\ast_{{\rm vs}}$ invariant. One shows that their algebra is trivial
$[C'[f],C'[g]]=0$ on this space. Note that there is no contradiction to the previous 
section because $C[f],\; C'[f]$ are operators defined on different spaces, or in other 
words, these are different representations. On both spaces 
the algebra closes but with an anomaly but it does not prevent the existence of non-trivial
solutions.

Note that habitat representations, in contrast to Hilbert space representations 
of $\mathfrak{h}$, do not come equipped 
with a Hilbert space structure or other topology. They are thus rather formal objects. If the constraints are not defined on the
kinematical HS, rigging methods to solve them and to provide a 
physical HS structure, are not available
and one could in fact have started with formal habitat representations
of the CCR without going through the exercise
to define HS  representations of the CCR and AR 
(implementation of the AR is impossible outside a Hilbert space context). In fact, how to close the constraint algebra formally using sufficiently differentiable 
functions has been shown already in \cite{60a,60b} for what one could now call the 
``loop representation habitat''. It was defined for Lorentzian self-dual gravity 
with density weight two but it applies verbatim 
to Euclidian gravity. One defined regulated operators by multiplying them with 
positive powers of $\epsilon$ (multiplicative renormalisation) and then taking limits.
The resulting algebra is formally closing without anomalies on the chosen habitat. To avoid confusion,
note that it is the kinematical HS structure that is needed to define the rigging map, not the operator topology defined above that was merely used to define 
the continuum limit of the Hamiltonian constraint on the 
kinematical HS.

\subsection{Solutions and propagation}
\label{s3.12}

In \cite{13a-2} a general framework was layed out for how to construct generalised (i.e.
distributional, elements of ${\cal D}^\ast$) solutions $l$ to all constraints, satisfying
$l[C[f]\;T]=l[(\tilde{U}(\varphi)-1)T]=0$ for all $f,\varphi,T\in {\cal D}$. Obviously 
the space of solutions ${\cal D}^\ast_{{\rm phys}}$ is a subspace of 
${\cal D}^\ast_{{\rm diff}}$ hence these are certain linear combinations of elements of 
${\cal D}^\ast_{{\rm diff}}$ and the Hamiltonian constraints imposes linear relations 
on the corresponding coefficients. In particular one can construct rather simple solutions 
consisting of linear combinations of diffeo inv. distributions labelled by a small
number of diffeomorphism classes of graphs, so that $l[C[f]T_{\gamma,j\iota}]=0$ is not automatically satisfied 
for a finite number of diffeo classes of $\gamma$ and by diffeo invariance we can 
restrict to one representative of each of those classes. The set of relations then
involves a countable set of coefficients labelled by spins, intertwiners and those diffeo classes
of graphs. 

The fact that $C[f]$ acts locally at a vertex 
suggests that the set of those relations can be solved for each vertex individually,
because we can restrict the support of $f$ to one of the vertices of $\gamma$. In other 
words the sets of relations obtained from the action at different vertices appear to
decouple, displaying {\it absence of propagation in the set of solutions} \cite{27} which 
would be physically inacceptable.

In \cite{28} it is argued (see below) that the decoupling of the sets of relations from different 
vertices {\it does not happen} generically. The confusion arises due to the following:
In \cite{13a-2} two versions of the Hamiltonian constraint were discussed, the one sketched in 
section \ref{s3.7} and \ref{s3.10} and another version which is more similar to the one discussed
in \cite{36,37} which much simplifies the structure of the set of solutions but which indeed 
leads to absence of propagation because of ``unique parentage'', see below. Thus the conclusion 
of \cite{27} applies to the second version, the analysis of \cite{28} to the first version.
  
The basic mechanism displayed in \cite{28} can be sketched as follows (we focus on the 
geometry contribution0:\\
The action of $C$ at a vertex $v$ of a SNWF over $\gamma$ results in a linear combination of 
SNWF over graphs $\gamma'$ differing from $\gamma$ in one (from $C^G_E$) or two
(from $C^G_L$) new, so called extraordinary edges $a$, between pre-existing ones $e,e'$ 
intersecting in $v$ such that $a$ intersects $e,e'$ transversally in interior points.
Consider a graph $\gamma$ without extra-ordinary edges with the property 
that three of its vertices $v_1,v_2,v_3$ are such that 
$v_1,v_2$ are joined by at least two edges $e_1,e_2$ and $v_2,v_3$ by at least two edges $f_1,f_2$. 
In particular, the action of $C$ at $v_1$ produces a  SNWF
over a graph $\gamma_1'$ with one new extra-ordinary edge $a_1$ between $e_1,e_2$. Likewise, the action at $v_3$ produces 
in particular a SNWF over a graph $\gamma_2'$ with one new edge $a_2$ between $f_1,f_2$. 
Finally the action at $v_2$ produces in particular a SNWF over $\tilde{\gamma}_1,\tilde{\gamma}_2$ 
respectively with extra-ordinary edges $\tilde{a}_1,\tilde{a}_2$ between $e_1,e_2$ and 
$f_1,f_2$ respectively. Consider now 
a solution $l$ to all constraints which consists of linear combinations of diffeomorphism
orbits of SNWF over graphs $\gamma'$ differing from $\gamma$ in one extra-ordinary edge
between $e_1,e_2$ or $f_1,f_2$. Then the equations $l[C[f] T_{\hat{\gamma},j,\iota}=0$ are automatically
satisfied unless $\hat{\gamma}$ is in the diffeo class of $\gamma$ in which case all equations 
are satisfied iff they are satisfied for $\hat{\gamma}=\gamma$ (by diffeomorphism invariance). 
We obtain 
three sets of equations coming from the action of $C$ at $v_1,v_2,v_3$ respectively. 
However, these equations are not independent of each other: The $v_1$ and $v_3$ equations 
are coupled by the $v_2$ equations because the graphs $\gamma'_I,\tilde{\gamma}_I; \; i=1,2$ 
are diffeomorphic.
If one just solves the $v_1, v_3$ equations respectively and specifies the respective free data
then those two sets of free data are brought into relation by the $v_2$ equations, which is a 
sense of propagation. This is the effect of non-unique ``parentage'' \cite{13d-1,13d-2,13d-3} which 
already has been there since \cite{13a-2}, i.e. 
modulo diffeomorphisms, a ``parent'' SNWF can be in the range, modulo diffeomorphisms, 
of the action of the Hamiltonian constraint from different vertices acting on different 
``child'' SNWF.     

In \cite{28} the simpler U(1)$^3$ model was considered because as compared to SU(2) for U(1)$^3$ 
one can determine the spectrum of the volume operator in closed form. Even in this case 
it is surprisingly difficult to make this mathematically water-tight (existence of solutions,
normalisablility wrt diffeo inv. scalar product) involving number theory and discrete PDE theory
but there is no reason to believe that the difference between U(1)$^3$ and SU(2) 
theory leads to absence of 
propagation.
 
\subsection{Interim summary: Anomalies and ambiguities}
\label{s3.13}

To summarise the developments so far: The Hamiltonian constraint $C[f]$ of \cite{13a-1a,13a-1b}
is densely defined on a Hilbert space $\cal H$ carrying a representation of the 
CCR (or CAR) and AR of geometry and matter. However, 1. it suffers from quantisation
ambiguities that survive the regulator removal limit $\epsilon\to 0$ 
such as those indicated in subsection \ref{s3.6} and 2. the constraint 
algebra generated by it and the spatial diffeomorphism constraint is anomalous in the sense 
that while the algebra still closes,
it closes with the wrong structure operators (quantisations of the structure functions).
It could have been worse: The commutators could have resulted not in linear combinations 
of Hamiltonian and spatial diffeomorphism constraints which would imply that the number 
of physical quantum degrees of freedom is lower than the number of physical classical degrees 
of freedom. However, the anomaly indicates that the quantum theory in its present form,
while not constraining the {\it wrong number} of degrees of freedom, selects the 
{\it qualitatively wrong} 
physical degrees of freedom. 

Clearly, the ambiguities and the anomaly must be adequately dealt with. These two 
problems are very likely linked to each other as one would expect that anomaly avoidance
decreases the amount of quantisation ambiguity. The developments that will be 
described below seek to avoid the anomaly in different ways. In the master constraint 
approach \cite{15,25a,25b,25c,25d} one avoids the algebra $\mathfrak{h}$ altogether by using a classically equivalent
single constraint. In the electric shift approach \cite{13d-1,13d-2,13d-3} one uses different density weights 
and habitats to construct a representation of $\mathfrak{h}$ in the sense of 
\cite{13e}. In the reduced phase space approach \cite{23a,23b,23c,23d,23e}, to which we devote a section of its 
own, the algebra $\mathfrak{h}$ is dealt with classically so that quantum anomalies 
cannot possibly arise. However, all
three approaches still suffer from quantisation ambiguities. Therefore, e.g. (Hamiltonian)
non-perturbative renormalisation \cite{13f} must be used as an additional step to remove 
the quantisation ambiguities.

Assuming that the anomalies can be properly removed,
compared to the perturbative QFT (or effective FT) approach to QG the progress 
of LQG to date can be stated as follows:\\
1.\\ 
In the EFT approach there are short distance (UV) infinities.
In LQG there are no UV infinities due to spatial diffeomorphism invariance.\\
2.\\
In the EFT approach, even after taming the UV infinities using perturbative renormalisation, 
there is a perturbation series to be summed with little control over the radius of 
convergence. In LQG there is no series to be performed because 
the approach is background independent and non-perturbative from the outset.\\
3.\\
In the EFT approach, perturbative renormalisation of the UV infinities requires adding new counter terms 
order by order making the theory non-predictive. In LQG there are no UV infinities and 
thus no counter terms required. However, there are quantisation ambiguities which 
also make the theory non-predictive so far. This stresses again the necessity to use 
non-perturbative renormalisation in LQG.

\subsection{(Extended) master constraint}
\label{s3.14}

The idea of the master constraint approach is quite simple: Given a set of classical
first class constraints $C_I$ possibly with structure functions rather than structure constants
consider instead the master constraint
\be \label{3.23}
{\bf\sf M}=\frac{1}{2}\;\sum_I\; C_I^\ast\; \omega_I\; \; C_J
\ee
(in the classical theory the constraints are real valued $C_I=C_I^\ast$) where $\omega_I>0$ 
are positive (weight) numbers. Then
\be \label{3.24}
C_I=0\;\forall I \;\;\Leftrightarrow\;\; {\bf\sf M}=0,\;\;\;
\{C_I, O\}_{{\bf\sf M}=0}\;\forall I\;\; \Leftrightarrow\;\; 
\{\{{\bf\sf M},O\},O\}_{{\bf\sf M}=0}=0
\ee
Thus {\bf\sf M} encodes not only the constraint surface but also the observables (reduced 
phase space). Hence, we consider constructing the single operator {\bf \sf M} rather 
than all the $C_I$. Since the quantisation of $C_I$ is however an integral part of 
quantising {\bf \sf M} the question arises how {\bf\sf M} encodes the anomaly and 
what influence the choice of $\omega_I$ has. To see this, suppose that zero is in the point 
spectrum of {\bf\sf M} and that ${\bf\sf M}\psi=0$. Then 
$<\psi,\; {\bf\sf M}\psi>=\sum_I\; \omega_I ||C_I\psi||^2=0$ thus $C_I\psi=0\;\forall \; I$
and vice versa. Thus, the presence of an anomaly in the $C_I$ will be encoded in the spectrum
of {\bf\sf M}. Removing the anomaly can then be considered as the problem to quantise
{\bf \sf M} such that it has a kernel at all or to minimise the lower bound of the spectrum 
of {\bf\sf M}. This proposal has been studied in various models \cite{17} where also 
the case of zero being part of the continuous spectrum or mixed cases was treated.  

For LQG the following concrete expression was quantised \cite{15,25a,25b,25c,25d}
\be \label{3.25}
{\bf\sf M}=\frac{1}{2}\int_\sigma\;d^3x\; \{\left[\frac{C}{Q^{1/2}}\right]^\ast\;\left[\frac{C}{Q^{1/2}}\right]
+\delta^{jk}\; \left[\frac{E^a_j\; D_a}{Q^{3/2}}\right]^\ast\; \left[\frac{E^b_k\; D_a}{Q^{3/2}}\right]\}
\ee 
The selection of the weight functions is motivated by spatial diffeomorphism invariance.
Note that not only the Hamiltonian constraint but also the spatial diffeomorphism
constraint is encoded in {\bf\sf M} (extended master constraint). 
This has three reasons: First, in contrast to 
$D_a$ itself, the function $D_j:=E^a_j D_a/Q$ or its ``square'' $q^{ab} D_a D_b/Q
=\delta^{jk}D_j D_k/Q$ 
{\it can be quantised} \cite{13a-3}. This kind of operator is also considered 
in the electric shift approach. Second, using the constraints $C$ and $D_j$ 
instead of $C,D_a$ makes their algebraic structure more alike. Finally, using the constraints 
$C,D_j$ we obtain a closed algebra with the advantage that in the quantum theory we 
can use $D_j$ itself rather than its exponentiation. There is also a version that 
just involves $C$ and which defines an operator on the Hilbert space extension of 
${\cal D}^\ast_{{\rm diff}}$ (non extended master constraint).

As the operator (\ref{3.25}) is spatially diffeomorphism invariant, it must act on the 
kinematical Hilbert space in a non-graph changing way. This can be done as follows:
Given a vertex $v$ of a graph $\gamma$ with two edges $e,e'$ outgoing from $v$ a loop
$\alpha$ in $\gamma$ based on $v,e,e'$ is called minimal iff it starts from $v$ along $e$, ends at 
$v$ along $(e')^{-1}$ and there is no other such loop with fewer edges traversed. 
The operator {\bf\sf M} uses such minimal loops instead of the extra-ordinary edges of $C[f]$ 
and projection operators making sure that the image of ${\bf\sf M}$ on a SNWF over $\gamma$
is a linear combination of SNWF over $\gamma$ and not a smaller graph. 
As {\bf\sf M} is graph preserving, its classical limit can be studied using the 
coherent state technology of \cite{24a,24b,24c,24d} and has been confirmed to converge to the classical 
expression plus quantum corrections for sufficiently large and fine graphs which justifies the above definition.

\subsection{Algebraic quantum gravity (AQG)}
\label{s3.14a}

As mentioned in the previous subsection, good semiclassical properties for {\bf\sf M}
are obtained for coherent states based on a single sufficiently large and fine graph. 
The non-separable kinematical Hilbert space is spanned by SNWF over finite graphs and there cannot be 
a semiclassical state on a single graph that controls the fluctuations of all quantum degrees of 
freedom, not even if we allow also graphs with a countably infinite number of edges 
(infinite tensor product (ITP) extension \cite{26}). On the other hand, it is clear 
that the kinematical Hilbert space is in some sense unnecessarily large: In the 
classical theory, e.g. a countable number of holonomies and fluxes would suffice to 
separate the points of the classical phase space. This has motivated the algebraic 
quantum gravity (AQG) viewpoint: One considers a countable number of quantum degrees 
of freedom that define an abstract $^\ast-$algebra $\mathfrak{A}$. The information of how to 
think of these algebra elements in terms of holonomies and fluxes along embedded paths 
and surfaces is part of the definition of a semiclassical state. In this way, the 
universe becomes a single abstract and infinite lattice $\lambda$ (i.e. only information which 
vertices are linked by which edges is provided). 

The corresponding Hilbert space representation of $\mathfrak{A}$ is formally the same 
as in LQG. However, now the spatial diffeomorphism constraint is no longer considered 
an extra structure but encoded in {\bf\sf M}. Spatial diffeomorphisms no longer 
act on $\lambda$ but just on $\mathfrak{A}$. The reason for why {\bf\sf M} had to 
be graph preserving on the LQG Hilbert space now no longer applies and {\bf\sf M} 
spreads its action unlimitedly over $\lambda$. The semiclassical analysis of the 
previous section still applies unchanged.

The AQG viewpoint rests on the selection of the abstract lattice $\lambda$. Using the 
huge freedom for how to embed $\lambda$ and since one can choose to leave some 
of its edges non-excited one can in fact accomodate an uncountably infinite number 
of what one would call embedded SNWF. Therefore AQG looks almost like a continuum 
theory which can be restricted to finite resolution as one desires and thus takes a large 
step towards non perturbative renormalisation. In fact in AQG quantisation ambiguities 
prevail even if one manages to minimise the spectral gap of {\bf \sf M} and one would prefer 
to get rid of the $\lambda$ dependence. 

\subsection{Renormalisation}
\label{s3.15}

This then gives direct motivation to enter non-perturbative (Hamiltonian) renormalisation (HR)
of LQG \cite{13f}. It applies simultaneously to the master constraint {\bf \sf M} 
of Dirac quantisation and to the physical Hamiltonian {\bf \sf H} 
of reduced phase space quantisation. As {\bf \sf H} is subject of the next section,
we focus here exemplarily on {\bf\sf M}.

The concrete proposal of \cite{13f} is motivated 
by constructive QFT \cite{49} but of course there were many earlier and related 
works, see \cite{13f} and references therein. In constructive QFT (CQFT) one works 
with a family of theories labelled by both IR and UV cut-off $R,\epsilon$ respectively. 
Then one first removes $\epsilon\to 0$ using the renormalisation flow and after that takes 
the thermodynamic limit $R\to \infty$. Consider the 3-torus $\sigma=T^3$ of radius $R$
and fields on $\sigma$ with periodic boundary conditions. We consider cubic lattices 
on $\sigma$ with $\epsilon^{-1}\in \mathbb{N}$ vertices in each direction. We define 
a partial order 
on the set ${\cal E}$ of resolutions $\epsilon$ 
by $\epsilon'\le \epsilon$ iff $\frac{\epsilon}{\epsilon'}\in \mathbb{N}$
is integral so that the coarser lattice is a sublattice of the former. The set $\cal E$
is also directed this way. Using some technology from wavelet theory 
\cite{50,51} (multi resolution analysis (MRA)) one constructs a one-particle Hilbert space
$V$ of smearing functions of the fields and finite resolution subspaces $V_\epsilon$ 
with $V_\epsilon\subset V_{\epsilon'},\; \epsilon'\le \epsilon$
thereof
together with embeddings $I_\epsilon:\; L_\epsilon\to V_\epsilon\subset V$ where 
$L_\epsilon$ is an $\ell_2$ space and $I_\epsilon$ is an isometry. This is all 
one needs to define the coarse graining map 
$I_{\epsilon\epsilon'}=I_{\epsilon'}^\dagger \; I_\epsilon:\; L_\epsilon\to L_{\epsilon'}$
and the projection $p_\epsilon=I_\epsilon I_\epsilon^\dagger:\; V\to V$. 

One considers the 
discretised fields $(\phi_\epsilon:=I_\epsilon^\dagger \phi,\;
\pi_\epsilon:=I_\epsilon^\dagger \pi)$ where $(\phi,\pi)$ is a collective notation for 
the continuum fields and discretised Weyl elements $w_\epsilon[F_\epsilon]=
w[I_\epsilon F_\epsilon]$ with $w[F]$ the continuum Weyl element which is a functional
of $<F,\phi>_V$. As an initial discretisation of {\bf\sf M} at resolution $\epsilon$
we pick ${\bf\sf M}^{(0)}_\epsilon[\phi_\epsilon,\pi_\epsilon]:=M[p_\epsilon \phi,p_\epsilon \pi]$.
The discretised fields are canonically conjugate if the continuum ones are and since 
in presence of both cut-offs the number of degrees of freedom is finite there is 
a unique Hilbert space representation $({\cal H}_\epsilon,\rho_\epsilon)$ of the 
discretised CCR and AR (Stone von Neumann theorem). We consider the ground state 
$\Omega^{(0)}_\epsilon\in {\cal H}_\epsilon$ and now construct a renormalisation flow 
or sequence of families of pairs $n\mapsto (\Omega^{(n)}_\epsilon,\;M^{(n)}_\epsilon)_{\epsilon\in {\cal E}}$ 
as follows: Pick a function $\kappa:\;{\cal E}\to {\cal E};\;\kappa(\epsilon)<\epsilon$
(often $\kappa(\epsilon)=\epsilon/2$) and define $\Omega^{(n+1)}_\epsilon$ such that 
\be \label{3.26}
J_{\epsilon\epsilon'} w_\epsilon[F_\epsilon]\Omega^{(n+1)}_\epsilon:=
w_{\epsilon'}[I_{\epsilon\epsilon'}\;F_\epsilon]\Omega^{(n)}_{\epsilon'}
\ee
is an isometry for $\epsilon'=\kappa(\epsilon)$ and set for $\epsilon'=\kappa(\epsilon)$
\be \label{3.27}
{\bf \sf M}^{(n+1)}_\epsilon:=J_{\epsilon\epsilon'}^\dagger\; {\bf\sf M}^{(n)}_{\epsilon'}\; J_{\epsilon\epsilon'}
\ee

At a fixed point of the flow (\ref{3.26}), (\ref{3.27}) we obtain the continuum Hilbert 
space $\cal H$ as the inductive limit of the ${\cal H}_\epsilon$, i.e. there exist 
isometries $J_\epsilon: {\cal H}_\epsilon \to {\cal H}$ with 
$J_{\epsilon\epsilon'}=J_{\epsilon'}^\dagger\; J_\epsilon$ for $\epsilon'=\kappa(\epsilon)$
and a continuum quadratic form {\bf\sf M} densely defined on the subspaces $J_\epsilon {\cal H}_\epsilon$
such that $J_\epsilon^\dagger \; {\bf\sf M}\; J_\epsilon={\bf \sf M}_\epsilon$ 
(``blocked from the continuum''). 

Although difficult to prove in general, one relies on universality and hopes that 
the inductive limit and quadratic form construction do not depend on the MRA and the map $\kappa$
which is true in the examples considered so far. The renormalisation flow has the tendency 
to reduce the number of ambiguities (couplings) to the ``relevant and marginal'' ones which yields 
a predictive theory if these are finite in number which is the whole point of renormalisation.
There is no guarantee that the positive quadratic form {\bf \sf M} extends to an operator
(if it does, one can take its Friedrichs s.a. extension). This programme is still
in its infancy and has been applied only to solvable QFT models so far.   

One can also apply the same flow equations to the constraints themselves rather than 
{\bf\sf M} in which case $\Omega_\epsilon$ is in general just a cyclic vector rather 
than a vacuum. The finite resolution algebra blocked from the continuum {\it must always 
be anomalous} even if the continuum algebra is non-anomalous, thus the ``finite resolution 
anomaly'', 
which should better be called a ``discretisation artefact'', is physically correct. 
See \cite{52} for an exactly solvable model 
and a technical explanation for the phenomenon.

\subsection{Electric shift approach}
\label{s3.16}

The electric shift approach has a longer history starting with first ideas tested in 
parametrised field theory \cite{53a-1,53a-2,53-1,53-2} then was generalised and improved in the U(1)$^3$ 
truncation of Euclidian vacuum GR \cite{54a,54b,54c} culminating recently in the 
treatment of full SU(2) Euclidian vacuum GR \cite{13d-1,13d-2,13d-3}. Common to these works 
is that one seeks a more geometrically motivated action of the Hamiltonian constraint
than chosen in \cite{13a-1a,13a-1b} which is more inspired by lattice gauge theory. In particular, one 
exploits the closeness of the generator of spatial diffeomorphisms and the Hamiltonian
constraint which {\it holds only for Euclidian vacuum GR}, i.e. only for the contribution
$C^G_E$ to the full Hamiltonian constraint. Therefore, the following exposition is 
strictly restricted to Euclidian vacuum GR. While there is possibly a chance to extend
this to the Lorentzian regime using the quantum Wick transform \cite{55,56} (see however
the reservations spelled out in \cite{13d-1,13d-2,13d-3}) the cosmological constant and matter 
contributions (and perhaps also the Lorentzian geometry contribution) are excluded. 
In fact these contributions are excluded for two reasons: First, because the close relation  
between diffeo constraint and Hamiltonian constraint does not extend to these contributions
and second because of the necessity to change the density weight.    

Namely, the electric shift approach considers a representation of $\mathfrak{h}$ on a certain
habitat, i.e. a certain subspace of the algebraic dual 
${\cal D}^\ast$, see subsection \ref{s3.11a} and 
\cite{13e}. To avoid the anomalous, Abelian character of the (dual) algebra discovered in 
\cite{13e}, in \cite{13d-1,13d-2,13d-3} one takes a drastic step: One changes the density weight away 
from unity. This means that a regulator limit on the kinematical Hilbert space or rather 
the dense and invariant domain $\cal D$ in the sense of
subsections \ref{s3.5}, \ref{s3.10} does not exist. This by itself may be
argued not to be problematic because one could simply not care about a representation 
of $\mathfrak{h}$ on $\cal D$ and is satisfied with a dual representation on a suitable 
habitat (see however the reservations at the end of subsection \ref{s3.11a}).
However, even to define the cosmological constant term with the modified 
density weight on that habitat is not possible \cite{57,13d-1,13d-2,13d-3}, see subsection \ref{s3.5}.
Work is in progress to return to density weight unity \cite{13d-1,13d-2,13d-3}. 

The reason for calling this the electric shift approach is the following: Recall that 
the density weight $w$ Hamiltonian and spatial diffeomorphism
constraint for Euclidian vacuum GR are respectively given by (modulo the SU(2) Gauss constraint)
\be \label{3.28}
C^G_E[f]=\int_\sigma\; d^3x\; \left[\frac{f\; E^a_j}{Q^{2-w}}\right]\; \left[\epsilon_{jkl} F_{ab}^k E^b_l\right],\;\;
D^G_a[u]=\int_\sigma\; d^3x\; u^a\; \left[F_{ab}^k E^b_k\right]    
\ee
where we have split the factors suggestively. Then $C^G_E[f]$ looks almost as a 
diffeomorphism with field dependent (electric) vector field (shift)
\be \label{3.30}
u^a_j(f)=\frac{f\; E^a_j}{Q^{2-w}}
\ee
proportional to $f$ (the lapse function). The rough idea is then to quantise $u^a_j$ independently 
from the rest of the constraint and to treat the rest along the lines of the spatial 
diffeomorphism constraint. In fact, in the U(1)$^3$ model the electric shift operator 
is diagonal in the analog of the SNWF basis so that $u^a_j(f)$ becomes basically a 
vector field depending on the SNWF labels and one can then truly proceed in analogy 
to the quantisation of the spatial diffeomorphism constraint. In the SU(2) theory this is 
less trivial but the basic idea is the same. 

The constraint algebra of the Euclidian vacuum Hamiltonian constraints at density weight $w$ 
is given by and again factorised suggestively
\be \label{3.31}
K^G_E(f,g):=\{C^G_E[f],C^G_E[g]\}=-\int_\sigma\; d^3x\; \left[F^k_{ab}\; E^c_k\right]\; 
\left[\frac{E^a_j E^b_l\delta^{jl}}{Q^{4-2w}} \omega_b(f,g)\right],\;\;\omega_b=f\;g_{,b}-g\;f_{,b}
\ee
This is again a spatial diffeomorphism along the electric field dependent vector field
\be \label{3.32}
v^a(f,g)=\left[\frac{E^a_j E^b_l\; \delta^{jl}}{Q^{4-2w}} \omega_b(f,g)\right]
\ee
As outlined in \cite{13a-5} it can be quantised on $\cal D$ at $w=1$  
see sections \ref{s3.10}, \ref{s3.11}, in particular (\ref{3.22}). Basically the first 
term becomes a local finite diffeomorphism along a vector field non-vanishing in the 
vicinity of a vertex that coincides with the 
tangent vector field on an edge while $\omega$ becomes a discrete derivative along another 
edge of the graph of a SNWF. This idea of using local, edge tangent field dependent diffeomorphisms 
first is described in \cite{13a-3} is also a central ingredient of the work \cite{13d-1,13d-2,13d-3} which 
there is applied also to $C^G_E$ itself. 

The choice of density weight taken in \cite{13d-1,13d-2,13d-3} is $w=\frac{4}{3}$. To see why note that 
then
\be \label{3.33}
u^a_j(f)=\frac{f[\epsilon^2 E^a_j]}{[\epsilon^3 Q]^{2/3}},\;
v^a(f,g)=\frac{[\epsilon^2 E^a_j][\epsilon^2 E^b_l]\delta^{jl}}{[\epsilon^3 Q]^{4/3}}
\ee
thus as envisaged they can be independently quantised using the technology of 
section \ref{s3.6}, contribute only at the vertices of a graph and will yield single and 
double sums respectively over its adjacent edges because $\epsilon^2 E$ becomes a flux operator.  
Then in a Riemann sum approximation of $C^G_E(f), \; K^G_E(f,g)$ we encounter the 
combination $\frac{1}{\epsilon}[F\epsilon^2]\cdot [E\epsilon^2]$. A similar 
combination appears in the spatial diffeomorphism constraint and can be approximated 
by something close to a finite diffeomorphism times the singular factor $1/\epsilon$
which prevents this object to have a limit as an operator on $\cal D$
with respect to the topology of section \ref{s3.10}. 

We consider now the habitat construction. The precise definition of that space is quite 
complicated, it distinguishes between ``non-degenerate'' vertices (at least tri-valent,
each triple of edges has non-coplanar tangents at the vertex) and more singular vertices 
(``kinks'') for whose treatment a Euclidian background metric $h$ and associated
Riemannn normal coordinates are employed, takes 
care of graph symmetries \cite{44}, the semi-analytic structure \cite{12a,12b}
and of the propagation heredity of section 
\ref{s3.12}. We therefore just sketch the main ideas, in particular we neglect the fact 
that the action of the constraint consists of a propagation part similar to \cite{13a-1a,13a-1b}
which does not play any role in the commutator calculation \cite{13e} and an electric 
diffeomorphism piece on which we focus solely,
oversimplify the formulas 
and refer the reader to the complete and rigorous case by case
analysis \cite{13d-1,13d-2,13d-3}. We mention that \cite{13d-1,13d-2,13d-3} uses the RS volume operator (which 
has a smaller kernel than the AL volume and in particular does not vanish on tri-valent kinks which are produced by the electric shift
generated diffeomorphisms)
and Tychonov regularisation which allow a more convenient expansion of SNWF into 
volume eigenfunctions as compared to using the AL volume operator and Poisson bracket
identities to treat inverse volume powers as it is employed in \cite{13a-1a,13a-1b}, 
see section \ref{s3.6},

Roughly speaking and not considering graphs with kinks,
a basis element of the habitat is a sum over SNWF 
with fixed spins and intertwiners over all graphs of a given diffeomorphism class $B$ 
without graph symmetries and kinks which has coefficients that depend on a function 
$F:\sigma^{|V(\gamma)|}\to \mathbb{R}$. Formally
\be \label{3.34}
l_{B,F;j,\iota}=\sum_{\gamma\in B}\; [F(\{v\}_{v\in V(\gamma)})]\;\;<T_{\gamma,j,\iota},\;>
\ee
which is very much like the proposal of \cite{13e} just that instead of the general  
function $F$ on the respective $V(\gamma)$ in \cite{13d-1,13d-2,13d-3}one considers 
the special function $F$ given by the product over vertices of the same function $f$. The subspace ${\cal D}^\ast_1$ of ${\cal D}^\ast$ defined by those special functions $F$ is not preserved by the constraint action, however, its range 
${\cal D}^\ast_2$ lies also 
in the domain of the constraint action so that commutators can be computed. We skip those details and consider now a 1-parameter 
group $\varphi^u_t$ of spatial diffeomorphisms generated by the vector field $u$ and 
the quantity $t^{-1}[\tilde{U}(\varphi^u_t)-1]$. As $\tilde{U}$ is not weakly continuous 
on $\cal H$, the limit $t\to 0$ cannot be computed there. On the other hand the dual action
gives (we drop the spin and intertwiner labels)
\be \label{3.35}
\tilde{U}'(\varphi^u_t)\;l_{B,F}
=l_{B,(\varphi^u_t)^\ast F}
\ee 
where the $\varphi(B)=B$ was used. Thus we can take the derivative with respect to $t$
provided that $F$ is at least $C^1$ and obtain 
\be \label{3.36}
i\;D'[u] l_{B,F}:=\left[\frac{d}{dt}\; \tilde{U}'(\varphi^u_t)\; l_{B,F}\right]_{t=0}
=l_{B,u[F]},\; u[F](\{v\}_{v\in V(\gamma)}):=\sum_{v\in V(\gamma)} u^a(v)\; 
\frac{\partial F}{\partial v^a}
\ee
This yields a representation of spatial diffeomorphisms on this habitat.

Turning to the regulated Euclidian vacuum Hamiltonian constraint $C^G_E[f]$, 
using the analogy with spatial diffeomorphisms, one quantises it such that 
in the action on a SNWF over some graph $\gamma_0$ one can essentially replace the 
electric shift (\ref{3.30})
by $f u^0$ where $u^0$ is a vector field defined by $\gamma_0$ 
(similar to (\ref{3.22}) with local 
support in the vicinity of vertcies (more precisely 
it is a sum of such terms)) and the remainder in analogy to a spatial 
diffeomorphism. Then one obtains, leaving out many details 
\be \label{3.37}
\left[(C^G_E[f])' l_{B,F}\right][T_{\gamma_0}]
:=\lim_{\epsilon\to 0}  
[(C^G_E[f])_\epsilon' l_{B,F}][T_{\gamma_0}]
=[D'[f u_0] l_{B,F}][T_{\gamma_0}]
=l_{B,\; (f\; u_0)[F]}[T_\gamma]
\ee
It follows for the commutator 
\be \label{3.38}
\left[\left[(C^G_E[f])',(C^G_E[g])'\right] l_{B,F}\right][T_{\gamma_0}]
=l_{B,[[f u^0,g u^0]\; u^0 F]}[T_\gamma]=l_{B,((f\;(u^0[g])-g\;(u^0[f])\;u^0)[F]}[T_\gamma]
\ee
On the other hand, for $K^G_E(f,g)$ one quantises it such that in the action 
on a SNWF over $\gamma_0$ the electric field dependent vector (\ref{3.32}) is 
essentially given by $(f\; u^0[g]-g\;u^0[f])\; u^0$ (more precisely it is a sum of 
such terms). Then indeed we obtain closure, again leaving out many details
\be \label{3.39}
[(K^G_E(f,g))'\; l_{B,F}][T_{\gamma_0}]
 :=\lim_{\epsilon\to 0}  
[(K^G_E(f,g))_\epsilon' l_{B,F}][T_{\gamma_0}]  
= [[(C^G_E[f])',(C^G_E[g])'] l_{B,F}][T_{\gamma_0}]
\ee
We close this subsection with a couple of remarks:\\
1. As mentioned, the close algebraic relation between $D^G_E, \; C^G_E$ was maximally exploited 
to arrive at this result. Comparing e.g. to the case of a (uncharged) scalar field 
$D^S_a=\pi\phi_{,a},\;2\;C^S=Q^{w-1}\;[\frac{\pi^2}{Q}+Q\;[q^{ab}\; \phi_{,a}\;\phi_{,b}+V(\phi)]$
it is no longer true that both $D^S$ and $C^S$ is linear in the ``curvature'' $\phi_{,a}$
as it is true for $D^G_E,\; C^G_E$ and which is essential for the above to work. 
Similar remarks hold for the YM and fermion (with mass terms) contributions . Furthermore
the density weight $w=4/3$ is also geared just to the $C^G_E$ contribution and does not 
work for the others.\\
2. There is a reduction of certain ambiguities because the quantisation of $C^G_E$ follows 
closely that of $D^G_E$. Thus $[\tilde{U}(\varphi_{v,e})-1]T_{\gamma_0}$ when expanded in 
terms of holonomies $g$ around the loops $\alpha_{v,e,e'}:=e\circ \varphi_{v,e}(e)^{-1}\circ 
\varphi_{v,e}(e') \circ e'^{-1}$ based at $v$ at which $e,e'$ are adjacent and 
where $\varphi_{v,e}$ is 
a diffeomorphism along a vector field with support close to $v$ and which on $e$ coincides
with the tangent of $e$ (in particular $\varphi_{v,e}(e)\subset e$, see (\ref{3.22})) gives an expression of 
the form ($\pi$ some spin representation) 
\begin{eqnarray}
 \label{3.40}
\pi(gh_{e'})-\pi(h_{e'})&=&[\pi(1_2+{\rm Tr}(g\tau_j)\tau_j+...)-1_\pi]\pi(h_{e'}) \nonumber\\
{\rm Tr}(g\tau_j) X^j_\pi \pi(h_{e'})+...
&=&{\rm Tr}(g\tau_j)\; \frac{d}{dt} \pi(e^{t\tau_j} \;h_{e'})+...\end{eqnarray}
thus uses the loop holonomy in the spin 1/2 representation and a right invariant 
vector field (quantisation of the flux operator). 
Thus it confirms the choice of spin 1/2 made in \cite{13a-1a,13a-1b} and removes some of the ambiguities 
pointed out in \cite{35}. On the other hand, inverse volume ambiguities and loop size 
type of ambiguities are still present. One also has to invoke 
an at least minimal amount of an additional type of freedom,  which in earlier works was suppressed 
by the principle of naturalness consisting e.g. in the treatment of kink contributions and addition of higher order terms (in $\epsilon$) which ensure both propagation and closure as discussed in \cite{13d-1,13d-2,13d-3}. 
Finally, the choice of a habitat itself 
is equivalent to a choice of representation and while the habitats do not come equipped 
with a (Hilbert space) topology, we expect that different choices of habitats lead 
to unitarily inequivalent representations of the algebra of observables
in the resulting physical Hilbert space obtained by supplying a Hilbert space structure
to the space of solutions. Thus, further reduction of ambiguities e.g. by renormalisation methods is 
still necessary.\\
3.\\
It is quite remarkable that for the first time one can make the details work out to close the algebra
without anomalies in a technically precise sense (since 
the construction is diffeomorphism covariant, also the dual commutators between spatial
diffeomorphisms and the Hamiltonian constraints work out as expected), in particular 
that in the actual sums involved the double sum of the commutator (\ref{3.38}) reduces 
to the single sums (\ref{3.39}) with exactly the right coefficients.
Note that in our simplified exposition
the action (\ref{3.37}) suggests that
all solutions to the spatial diffeomorphism constraint with no kinks and graph symmetries (corresponding to constant
$F$) are already in the kernel of $(C^G_E[f])'$. However, the proper
treatment establishes that those states are not even in the domain
of the dual Hamiltonian constraint. Thus, in agreement with 
one's intuition, the two constraints have individual solutions not in the joint kernel.\\
4. \\
An interesting observation is the following \cite{58}: Since the action of $C^G_E$ is so close 
to a spatial diffeomorphism, one may wonder whether it can be supplied with a 
corresponding geometrical
interpretation. This is indeed the case: Note first that $D^G_E$ 
generates ordinary Lie derivatives ${\cal L}_u$ along ordinary vector fields $u$. These 
derivatives are not Gauss gauge covariant when acting on Lie algebra valued fields. 
However, the combination 
$\hat{D}^G_E=D^G_E+A^j_\cdot G_j$ used in (\ref{3.28}) does and generates gauge covariant Lie derivatives 
$\widehat{{\cal L}}_u$. Likewise, the Euclidian vacuum 
Hamiltonian constraint generates generalised gauge covariant Lie derivatives 
$\widehat{\widehat{{\cal L}}}_{\vec{u}}$ where now $\vec{u}^a_j$ is a Lie algebra valued 
vector field (namely the electric shift). The definition of these Lie derivatives 
differs from the ordinary one by replacing $\partial$ in the formula for ${\cal L}$ 
by $\cal D=\partial+A$ when $\partial$ would act on a Lie algebra valued field in order 
to maintain gauge covariance. These derivatives can be extended to 
vector fields of any internal tensor degree forming a generalised 
(open) algebra. Note however that $u^a_j$ is phase space dependent and not a test field. This geometric interpretation supplies further motivation to quantise
$C^G_E$ in close analogy to $D^G_E$. See also the next subsection
for a concrete quantisation of this algebra in the U(1)$^3$ model.

\subsection{Quantum non degeneracy}
\label{s3.17}

The theory laid out so far has revealed that to arrive at a representation of $\mathfrak{h}$
one has to take non-standard steps, either by looking at non-standard operator topologies 
when an implementation on $\cal D$ is intended as sketched in section \ref{s3.10} or by using 
non-standard density weights when an implementation on subspaces (habitats) of ${\cal D}^\ast$
is intended as sketched in the previous subsection. One may wonder how these complications 
arise and how non-standard constructions can be avoided. 

We note that in both implementations the density weight $w$ is lower than two. Therefore in 
the classical theory we have $\{C[f],C[g]\}=-
D_a[Q^{w-2} E^a_j E^b_k \delta^{jk}(f g_{,b}-g f_{,b})]$ depending on a negative power 
of $Q=|\det(E)|^{1/2}$. It is an implicit assumption 
of {\it classical} GR that $Q$ is nowhere vanishing, i.e. {\it non-degenerate}. We may consider  
a {\it relaxation} of this assumption in {\it quantum} GR but we then expect problems in constructing a 
representation of $\mathfrak{h}$ due to the negative power of $Q$. In fact the dense 
domain $\cal D$ of $\cal H$ defined by the span of SNWF is such that the volume operator 
$V(R)=\int_R \; d^3x\; Q$ vanishes on any element of $\cal D$ unless $R$ intersects 
a Lebesgue measure zero set. In that sense the quantum geometries described by 
SNWF are {\it quantum degenerate almost everywhere}. It is for this reason that 
one had to quantise inverse powers of $Q$ carefully as it was sketched in subsection
\ref{s3.6} with the result that inverse powers of $V(R)$ annihilate a SNWF unless 
$R$ contains at least one non-zero volume vertex of the underlying graph. Still this 
brings us into the awkward situation that we try to implement the constraints 
on a quantum domain whose semiclassical limit {\it is classically forbidden and on 
which the classical $\mathfrak{h}$ is ill-defined}. 

This explains why one is forced into the above non-standard steps as follows:
The classical constraints $D[u], C[f]$ are integrals over $\sigma$ of the densities 
$D_a(x)$ and $C(x)$ respectively and in order to define the operator corresponding 
to $C[f]$ we used a standard point-splitting regularisation based on a Riemann sum 
approximation $C_\epsilon[f]=\sum_\Delta\; f(p_\Delta) \; C_\Delta$ with 
$C_\Delta=[\int_\Delta\; d^3x\; C(x)]$. Here $p_\Delta\in \Delta$, 
the cells $\Delta$ have coordinate volume $\epsilon^3$ and there are an 
order of $N_\epsilon:=\epsilon^{-3}$ terms if $\sigma$ is compact (in the non-compact case, consider
$f$ of compact support) which makes sure that the sum converges to a non-zero limit 
(the integral) as $\epsilon\to 0$. Now in the quantum theory we quantise $C_\Delta$ on a 
SNWF $T_\gamma$ and eventually at most $N_\gamma:=|V(\gamma)|$ cells contribute as 
$\epsilon\to 0$. Due to diffeomorphism invariance, the norm $||C_\epsilon[f] T_\gamma||$ is 
finite, non-vanishing and $\epsilon$ independent if we define $p_\Delta=v$ when 
$\Delta \cap V(\gamma)=\{v\}$. However, for the analogous quantisation $K(f,g)$ of the Poisson 
bracket $\{C(f),C(g)\}$ sketched in section \ref{s3.11} the limit of the 
norm vanishes for $w=1$ unless $f,g$ have discontinuities. This is because $K_\epsilon(f,g)$ again 
has an order of $N_\gamma$ contributions but now depends on the finite differences schematically 
denoted as $f(v) g(v+\epsilon \dot{e}(v))-g(v) f(v+\epsilon \dot{e}(v))$ where $e$ is an 
edge of $\gamma$ adjacent to $v$. This is also the underlying reason why the algebra 
on the habitat \cite{13e} is Abelian and why in the previous section a different 
density weight was considered so that instead the combination 
$[f(v) g(v+\epsilon \dot{e}(v))-g(v) f(v+\epsilon \dot{e}(v))]/\epsilon$ results.

Thus while in the classical theory $K(f,g)$ is always non-vanishing for smooth $f,g$ and any 
density weight in the quantum theory we obtain this peculiar and awkward behaviour. The 
source of the trouble is the {\it quantum degeneracy} of the domain $\cal D$ or 
the entire representation. If one would work in a representation which is 
{\it quantum non-degenerate} then the number of contributing $C_\Delta$ would be 
an increasing function of $\epsilon^{-1}$ and there would be a chance that 
this awkward behaviour can be avoided.

In \cite{42a,42b,42c,42d} a {\it quantum non-degenerate 
representation} for U(1)$^3$ quantum gravity (which is a toy model very close to 
Euclidian vacuum GR and presents a consistent deformation in terms of a small 
Newton constant of 
Euclidian vacuum GR in the sense of \cite{61}) was found which allows for an anomaly 
free representation of the {\it Bergmann-Komar ``group''} $\mathfrak{H}:=\exp(\mathfrak{h})$ 
(the exponentiation avoids different treatments of finite diffeomorphisms and infinitesimal 
Hamiltonian constraints; the BK group is here defined as the Lie 
group defined by the true Lie algebra obtained by taking commutators of all hypersurface deformations (universal enveloping algebra of \cite{5})) for any density weight, on the corresponding $\cal D$ without using habitats. 
The model is {\it quantum integrable}, i.e. a physical Hilbert space representation can be found, because in contrast to the SU(2) theory the U(1)$^3$
has the property that the Hamiltonian constraint preserves the momentum polarisation of 
the phase space in the sense of geometric quantisation \cite{59}. This can be considered 
as a quantisation of the generalised gauge covariant derivatives of \cite{58} (see previous section)
{\it to all orders}. The model can be extended by a 
cosmological constant term. An intriguing idea is that one could perhaps define the full SU(2) theory
in terms of perturbation theory around this integrable theory as spelled out in \cite{61}. Note that while 
the standard LQG representation and the non-degenerate 
representation for U(1)$^3$ are different, they are still very similar in the sense that they use Narnhofer-Thirring type repesentations and thus much of the 
technology developed for LQG can be tranferred. In particular, all the results for U(1)$^3$ were obtained 
by exactly the same steps as in the present LQG 
representations with only minor modifications due to 
the different gauge groups in place.

This completely solvable model \cite{42d}, which is quantised following step by step the full arsenal of 
technologies developed for LQG,
may serve as a proof of principle that {\it making quantum degeneracy a prerequisite for 
representing $\mathfrak{h}$ or $\mathfrak{H}$} maybe a promising direction to make progress as spelled out in detail in \cite{13g}.
To arrive at such representations systematically or constructively, renormalisation 
methods suggest themselves because as sketched in section \ref{s3.15} in the renormalisation 
programme one works, at each finite resolution, by definition with a dense set of 
non-degenerate states for that resolution and thus at infinite resolution one expects 
quantum non-degeneracy to be inherited.

\section{Constraining before quantisation}
\label{s4}
In the reduced approach GR has been cast into the framework of an ordinary Hamiltonian theory. Therefore, the quanitsation requires to find representations of the observable algebra and the quantum dynamics can be implemented by quantising the physical Hamiltonian. It is also possible to perform only a partial reduction typically with respect to the Hamiltonian constraint only and then solve the remaining constraint via Dirac quantisation. For instance in the existing models \cite{23a,23b,23c,23d,23e} the Gau\ss{} constraint is solved using Dirac quantisation and either one or four respectively dust or scalar field are used as reference fields for the Hamiltonian and spatial diffeomorphism constraint respectively. A classification of the different type of models can be found in \cite{Giesel:2012rb}, where a generic Lagrangian was analysed that for appropriate choices of the involved parameters encodes the existing dust models in the literature
 \cite{21-1,21-2,nulldust} based on seminal work of Kucha\v{r} et. al. These kind of models, denoted as type I models in \cite{Giesel:2012rb}, all have in common that one couples six to eight additional fields to gravity yielding to a second class system. Performing a symplectic reduction with respect to the second class constraints leads to a system that involves 4 (3) additional (null) dust fields that can, in the case of dust, be used as a dynamical reference frame and for null dust as a dynamical spatial reference frame. The remaining Hamiltonian, spatial diffeomorphism  and Gau\ss{} constraints are first class so that one can apply the usual construction of observables in the framework of the relational formalism in these models. A further type I model not involved in \cite{Giesel:2012rb} can be found in \cite{23e} where one Klein-Gordon scalar field as well as 6 additional scalar fields were minimally coupled to GR. The second class of models, denoted as type II in \cite{Giesel:2012rb}, involves a matter Lagrangian based on one scalar field and hence within such models only a partial reduction of the classical constraints can be achieved. Examples for such models can be found in \cite{23b,23f}.  As mentioned above the choice of reference matter is strongly guided by the aim to obtain a manageable observable algebra in the reduced models. As we will discuss below for all these models the reduced quantisation programme can be completed and the final physical Hilbert space is known. Because the reduced models always involve a true Hamiltonian that is non-vanishing they are also of advantage if one wants to generalise to open quantum systems were one often starts with a given Hamiltonian of the total system. First steps in this direction using the relational formalism and a reduced quantisation  can be found for instance in \cite{Fahn:2022zql}.

\subsection{Reduced quantisation of type I models}
\label{s4.1}
In the used notation $(Q^A,P_A)$ collectively denote all degrees of freedom not related to the reference fields and $(T^I,P_I)$ those of the reference fields, where for type I models $I=0,\cdots,3$ and $A$ labels the remaining degrees of freedom.  All variables $Q^A,P_A$ that we want to construct observables of Poisson commute with the chosen reference fields $T^I$. The corresponding gauge fixing conditions $F_I=T^I-\tau^I$ being linearly in these reference fields also commute with the remaining variables and consequently agrees the Dirac bracket with the Poisson bracket in this case. The observable algebra of $O_{Q^A}, O_{P_A}$ thus reads
\begin{equation*}
\{ O_{Q^A}(\sigma,\tau^0),O_{P_B}(\sigma^\prime,\tau^0)\}=\lambda\delta^A_B\delta(\sigma,
\sigma^\prime),
\end{equation*}
where $\tau^I=(\tau^0,\sigma^j),\, j=1,2,3$  denote the physical temporal and spatial coordinates, $\lambda$ is a possible coupling parameter that can also be equal to 1 and all remaining Poisson brackets vanish so that the algebra of these observables satisfies standard CAR and AR. Since for type I models $O_{Q^A}, O_{P_A}$ are observables with respect to the Hamiltonian and spatial diffeomorphism constraints a representation of the observable algebra yields direct access to the physical Hilbert space if one solves the Gau\ss{} constraint in the quantum theory.
Next to solving the Gau\ss{} constraint one is only interested in those representations for which the physical Hamiltonians of the individual models can be promoted to well-defined operators. Three examples for physical Hamiltonians $H$ can be found in \eqref{eq:Hphys} below
\begin{eqnarray}
\label{eq:Hphys}
H & = &\int\limits_{\cal S} d^3\sigma H(\sigma) \\
H(\sigma) &=& \sqrt{(O_C)^2 -Q^{jk}O_{D_j}O_{D_k}}(\sigma)\quad 
\mbox{(Brown-Kucha\v{r})}\quad  [29] \nonumber \\
H(\sigma) &=& O_C(\sigma) \quad (Gaussian) \quad [59] \nonumber \\
H(\sigma) &=& \sqrt{-2\det(Q)O_C + 2\sqrt{\det(Q)\sum\limits_{j=1}^3 Q^{jj}O_{D_j}O_{D_j}}},\quad (4\, scalar\, fields) \quad [60] \nonumber
\end{eqnarray}
where ${\cal S}$ symbolises the manifold of physical spatial coordinates $\sigma^j$ and  $Q_{jk}:=O_{q_{jk}}$ denotes the observable of the spatial metric understood as a function of the densitised (co-)triads $E$. $O_C$ and $O_{D_j}$ respectively denote the observable of the contribution to Hamiltonian and spatial diffeomorphism constraint respectively of the physical sector encoded in $O_{Q^A},O_{P_A}$. As one can see a generic feature of all these models is that the physical Hamiltonian densities $H(\sigma)$ often involve square roots and inside the square usually powers of $O_{C}$ and $O_{D_j}$ together with contractions with the metric as well possible density weight factors occur, where the latter ensure that $H(\sigma)$ is a scalar density of weight one. 
 Due to the simple structure of the observable algebra one can use the standard LQG representation that was used in section \ref{s3} for the kinematical Hilbert space here for the physical Hilbert space if one considers in addition SU(2) gauge invariant SNWF. Although an operator for $O_{C_j}$ does not exist in this representation for the physical Hamiltonians one only needs to be able to quantise the combination $Q^{jk}O_{D_j}O_{D_k}$ which, as will be explained below, is possible to quantise along the lines how the operator for $O_C$ is constructed. Note that not exactly the same but a similar expression is also involved in the extended master constraint in section \ref{s3.14}. Written as a function of the original kinematical variables $Q^A,P_A,T^I,P_I$ the observables $O_C,O_{D_j}$ are complicated because already the elementary observables $O_{Q^A},O_{P_A}$ are in general an infinite power series in the reference fields with phase space dependent coefficients. However, one can use the properties of the observable map to show that \cite{22a,22b,22c,22d,22e}
 \begin{equation*}
O_{C}\simeq C(O_{Q^A},O_{P_A})\quad O_{D_j}\simeq D_j(O_{Q^A},O_{P_A}),
 \end{equation*}
where $\simeq$ denotes a weak equality. Therefore, one can apply  the strategy how $C$ was quantised, discussed in section \ref{s3.7}, also here and quantise $(O_{C})^2$ as $\hat{O}_C^\dagger\hat{O}_C$ using similar steps. The corresponding formula to \eqref{3.28} involving the Euclidian part of $C$ in the reduced case is given by
\begin{eqnarray} \label{eq:OC}
(O_{C^G_E})^2(\sigma)&=&\frac{\; \epsilon^{JKL}\, (O_{F})^J_{k\ell} (O_{E})^k_K (O_{E})^\ell_L\epsilon^{J'K'L'}\, (O_{F})^{J'}_{k'\ell'} (O_{E})^{k'}_{K'} (O_{E})^{\ell'}_{L'}}{\det(O_E)}(\sigma) \\
&=& \left[{\rm Tr}(B\tau_0)\right]^2(\sigma),\quad B:=B^j_{J'} e^J_j\tau_{J'}\tau_{J}=\frac{1}{2}\epsilon^{jk\ell}(O_F)^{J'}_{k\ell}e^J_j\tau_{J'}\tau_J\nonumber
\end{eqnarray}
where $\tau_\mu:=(\tau_0:=\mathds{1},\tau_I=-i\sigma_I)$, with $\sigma_I$ being the Pauli matrices,  $e^J_j$ denotes the co-triad and indices referring to physical coordinates are labeled by lower case later and those referring to internal SU(2) indices by capital letters. In the explicit action of the corresponding operators the contribution of the curvature $O_F$ involved in $B$ is thus quantised as a loop of holonomy operators that goes along already existing edges of the underlying graph that the SNWF is defined on. For the reason that the representation of the holonomy operator can couple with representations associated to the edges to the trivial representation and this would correspond to an edge that is annihilated, one needs to add corresponding projection operators to the physical Hamiltonian operator $\hat{H}$ in order ensure that it is indeed graph-preserving \cite{23a}. The contribution of $Q^{jk}O_{D_j}O_{D_k}$ in terms of the $A,E$ variables reads
\begin{eqnarray}
\label{eq:QDD}
Q^{jk}O_{D_j}O_{D_k}(\sigma) &=& 
\frac{(O_{E})^j_J (O_{E})^k_K(O_F)^{J'}_{j\ell}(O_E)^\ell_{J'}(O_F)^{K'}_{k\ell'}(O_E)^{\ell'}_{K'}\delta^{JK}}{\det(O_E)}(\sigma) \\
&=&\frac{1}{4}\left[{\rm Tr}(B\tau_I)\right]\left[{\rm Tr}(B\tau_J)\right]\delta^{IJ}.
\end{eqnarray}
Therefore, the basic building block of the individual physical Hamiltonians is ${\rm Tr}(B\tau_\mu)$ and certain powers thereof respectively. The quantisation of ${\rm Tr}(B\tau_\mu)$ can be performed along the lines of how the Hamiltonian constraint is implemented in the context of the Dirac quantisation. However, there exist one difference compared to case of the Dirac quantisation where the constraints but not a physical Hamiltonian is quantised. At the classical level the physical Hamiltonian is a spatially diffeomorphism invariant quantity and one aims at defining the corresponding operator also with the same symmetry. As has been shown in \cite{44} spatially diffeomorphism invariant operators that are graph-modifying do not exist in the LQG representation. Therefore, one needs to quantise the physical Hamiltonians in a graph-preserving manner. The physical Hilbert space based on the LQG representation involves SNWF defined on all graphs that can be embedded in the spatial manifold. Thus, graph-preserving or also called graph non-changing operators always come along with an infinite number of conservation laws, one for each graph, that are completely absent in the classical theory. Although such a graph-preserving property might be of advantage if one want to use current semiclassical techniques as discussed already in section \ref{s3.14}, here an additional motivation to work in the AQG framework is to avoid these additional conservation laws in the quantum theory. Because all physical Hamiltonians are spatially diffeomrphism invariant by construction they can be promoted to operators using the ITP Hilbert space AQG is based on and the usual quantisation strategy for these operators. 
For the models displayed in \eqref{eq:Hphys} the usual embedded LQG as well as their corresponding AQG quantisation exist \cite{23a,23e,Giesel:2012rb}. As can be seen in \eqref{eq:Hphys} the contribution related to the spatial diffeomorphisms enters differently into the Hamiltonian density of the 4 scalar field model but one can nevertheless quantise this quantity in the LQG and AQG representation as shown in \cite{23e}.

\subsection{Reduced quantisation of type II models}
Examples for quantum models of type II can be found in \cite{23b,23c,23f} that use either a massless scalar field, non-rotational dust and a phantom field often used in k-essence respectively as a reference field for the Hamiltonian constraint.  The corresponding physical Hamiltonians of these model are given by
\begin{eqnarray}
{H}&=&\int\limits_\sigma d^3 x\, \sqrt{-\sqrt{\tilde{Q}} \widetilde{O}_C+\sqrt{\widetilde{Q}} \sqrt{\left(\widetilde{O}_C\right)^2-\widetilde{Q}^{ab} (\widetilde{O}_D)_a(\widetilde{O}_D)_b}}, \quad [41] \\
{H} &=&\int\limits_\sigma d^3 x\, {\rm sgn}(\widetilde{O}_C)\widetilde{O}_C\quad [71]\nonumber \\
{H} &=&\int\limits_\sigma d^3 x\, \sqrt{\frac{1}{2}\left[(\widetilde{O}_C)^2-\widetilde{O}_{qDD}-\alpha^2 \widetilde{Q}\right]+\sqrt{\frac{1}{4}\left[(\widetilde{O}_C)^2-\widetilde{O}_{qDD}-\alpha^2 \widetilde{Q}\right]^2-\alpha^2  \widetilde{O}_{qDD}Q}}\nonumber  \\
{\rm with} &&  \widetilde{O}_{qDD}:=\widetilde{Q}^{ab} (\widetilde{O}_D)_a(\widetilde{O}_D)_b, \quad [119]
\nonumber 
\end{eqnarray}
where $\alpha$ is a parameter involved in the Lagrangian of the model in \cite{23f} and a tilde is used on the top of all observables because these observables are only constructed with respect to the Hamiltonian constraint. Hence, the main difference to the type I models is that here the spatial diffeomorphism constraint needs to be solved in the quantum theory. As a consequence, the usual LQG representation cannot be used for the physical Hilbert space here. This has been circumvented in \cite{23c} by using the diffeomorphism invariant Hilbert space ${\cal H}_{\rm diff}$, usually constructed in the Dirac quantisation approach,  as the physical Hilbert space. An alternative is to work in the framework of AQG where these physical Hamiltonians can be quantised using the standard procedure for the individual contributions inside the (double) square roots following the strategy discussed in section \ref{s4.1}. The model in \cite{23f} was discussed at the classical level only so far in the literature but can in principle be quantised with the same techniques. Note that if one works on ${\cal H}_{\rm diff}$ then one expects that the operator $\widehat{\widetilde{O}}_{qDD}$ annihilates all states in this space, therefore these contributions were neglected in \cite{23b} by hand. In this case one has $H=\int d^3 x \sqrt{-2 \sqrt{\tilde{Q}(x) \widetilde{O}_C}}(x)$ and this model can be understood as the full GR generalisation of the APS model \cite{APS-1,APS-2,APS-3} in LQC where the inflaton is chosen as the clock and when the contribution of the potential is subdominant. In case one fully incorporates the potential in both models than the final physical Hamiltonian becomes time-dependent and the usual issues of finding a suitable physical inner product in such situations is present. This can be avoided by coupling the reference matter in addition to the inflaton potential as has for instance be done in \cite{21a-1,21a-2,LQCclocks} with the price to pay that in the cosmological context one needs to work with two-fluid models. As a first attempt in \cite{23e} a coupling of 4 Klein-Gordon scalar fields as reference fields was considered in order to define a corresponding type I model of the type II model in \cite{23b}. However, as shown in \cite{23e} such a model yields to a physical Hamiltonian that does contain the contribution of $O_{D_j}$ not in the combination $Q^{jk}O_{D_j}O_{D_k}$ or individual  components thereof but involves  $\delta^{jk}O_{D_j}O_{D_k}$. As the discussion above showed in order to quantise these quantities in the LQG representation the contraction with the inverse metric is crucial as otherwise the operator does not exist. Thus, if one restricts to the LQG representation then for the model in \cite{23b} based on the Dirac quantisation of the spatial diffeomorphism the quantisation programme can be completed whereas this is not possible by simply using ordinary Klein-Gordon scalar fields as reference fields in the reduced case. The latter issue can be avoided by adding three more fields to the system in a suitable manner so that the system also becomes second class. A reduction with respect to the second class constraints yields a first class system with the physical Hamiltonian shown in \eqref{eq:Hphys} for which the quantisation programme can be similarly completed. In this sense a comparison of type I and type II models provides also some insights on the similarities and differences between Dirac and reduced quantisation as far as the spatial diffeomorphisms are concerned. This, however, requires, if not working in the full theory, to consider symmetry reduced models where $O_{D_j}$ does not trivially vanish, which goes beyond  homogeneous and isotropic models.

\section{Summary and outlook}
\label{s5}

Our exposition of the status of the dynamics of LQG suggests the following:\\
1.\\
Despite some progress, the quantisation before constraining route is still far away 
from making contact to phenomenology. Even after the issue of anomalies has been
settled in the full theory for Lorentzian signature including all observed matter,
one would still need to compute the kernel of the constraints, equip it with an inner product
and construct operator representations of Dirac observables thereon.
On the other hand, the techniques developed in this route are employed almost without change in the reduced phase space approach 
and are thus of outmost importance.\\
2.\\
The quantisation after constraining route in that sense is much more economic and sidesteps 
all of these complications. While to be practically useful it requires to use matter to
extract the reduced phase space, this is not a disadvantage in view of the fact that 
a universe without matter is unphysical, thus one may as well use it to make progress and 
also to move much closer to phenomenology.\\
3.\\
There is one technical issue common to both routes, which is the presence 
of quantisation ambiguities. These have to be downsized to a finite number to render 
the theory predictive. Renormalisation is a possible avenue to reach this goal.

\bibliography{REfQGHB}
\bibliographystyle{plainnat}
\end{document}